\documentclass[12pt]{article}

\usepackage[superscript,nomove]{cite}

\usepackage{times}
\usepackage{graphicx}
\usepackage{lineno}
\usepackage{longtable}
\usepackage{booktabs}
\usepackage{url}

\usepackage[margin=0.75in]{geometry}
\usepackage{fancyhdr}
\usepackage{hyperref}
\newcommand{\ion}[2]{#1~\RomanNumeralCaps{#2}}

\newcommand{\kms}{km~s\ensuremath{^{-1}}}
\newcommand{\vlsr}{\ensuremath{v_\mathrm{LSR}}}

\newcommand{\RomanNumeralCaps}[1]
    {\MakeUppercase{\romannumeral #1}}

\title{Observations of a Magellanic Corona}

\author
{Dhanesh Krishnarao,$^{1,2,3\ast}$ Andrew J. Fox,$^{4}$ Elena D'Onghia,$^{5}$\\
Bart P. Wakker,$^{5}$ Frances H. Cashman,$^{1}$ J. Christopher Howk,$^{6}$\\
Scott Lucchini,$^{7}$ David M. French,$^{1}$ Nicolas Lehner$^{6}$\\ 
\\
\normalsize{$^{1}$Space Telescope Science Institute}\\
\normalsize{3700 San Martin Dr, Baltimore, MD 21218, USA}\\
\normalsize{$^{2}$William H. Miller III Department of Physics \& Astronomy, Johns Hopkins University}\\
\normalsize{3400 N. Charles Street, Baltimore, MD 21218, USA}\\
\normalsize{$^{3}$Department of Physics, Colorado College}\\
\normalsize{14 East Cache La Poudre Street, Colorado Springs, CO 80903, USA}\\
\normalsize{$^{4}$AURA for ESA, Space Telescope Science Institute}\\
\normalsize{3700 San Martin Dr, Baltimore, MD 21218, USA}\\
\normalsize{$^{5}$Department of Astronomy, University of Wisconsin-Madison}\\
\normalsize{475 N Charter St, Madison, WI 53706, USA}\\
\normalsize{$^{6}$Department of Physics \& Astronomy, University of Notre Dame}\\
\normalsize{225 Nieuwland Science Hall, Notre Dame, IN 46556, USA}\\
\normalsize{$^{7}$Department of Physics, University of Wisconsin-Madison}\\
\normalsize{475 N Charter St, Madison, WI 53706, USA}\\

\\
\normalsize{$^\ast$To whom correspondence should be addressed; E-mail:  dkrishnarao@coloradocollege.edu.}
}

\date{}

\begin{document} 


\baselineskip24pt


\maketitle

{\bf The Large and Small Magellanic Clouds (LMC/SMC) are the closest major satellite galaxies of the Milky Way. They are likely on their first passage on an infalling orbit towards our Galaxy \cite{Besla2007}
and trace the ongoing dynamics of the Local Group \cite{D'Onghia2016}.
Recent measurements of a high mass for the LMC 
($M_{\rm halo}\approx10^{11.1-11.4}\,M_\odot$) \cite{Penarrubia2016,Erkal2018,Erkal2019,Kallivayalil2018} 
imply the LMC 
should host a Magellanic Corona: a collisionally ionized, warm-hot gaseous halo
at the virial temperature ($10^{5.3 - 5.5}~K$) initially extending out to the virial radius ($100 - 130$ kpc).
Such a Corona would have shaped the formation of the Magellanic Stream \cite{Lucchini2020}, a tidal gas structure extending over $200^\circ$ across the sky \cite{Besla2012,Nidever2010,D'Onghia2016} that is bringing in metal poor gas to the Milky Way \cite{Fox2014}. 
No observational evidence for such an extended Corona has been published previously, 
with detections of highly ionized gas only reported in directions directly toward the LMC, where winds from the LMC disk may dominate \cite{deBoer1980,Wakker1998}
Here we show evidence for this Magellanic Corona with a potential direct detection in highly ionized oxygen (O$^{+5}$), and indirectly via triply-ionized carbon and silicon, seen in ultraviolet absorption toward background quasars. 
We find that the Magellanic Corona is part of a pervasive multiphase Magellanic circumgalactic medium (CGM) seen in many ionization states with a declining projected radial profile out to at least $35$~kpc from the LMC and a total
ionized CGM mass of $\log_{10}(M_\mathrm{\ion{H}{2},CGM}/M_\odot) \approx 9.1 \pm 0.2$. 
The evidence for the Magellanic Corona is a crucial step forward in characterizing the Magellanic Group and its nested evolution with the Local Group.
}

We use a sample of 28 Hubble Space Telescope (\emph{HST})/Cosmic Origins Spectrograph (COS) spectra of background UV-bright quasars within an angular separation of the LMC of 45\textdegree, corresponding to an impact parameter $\rho_\mathrm{LMC} < 35~\mathrm{kpc}$, spanning one-third of the LMC's initial virial radius. 6 of these sightlines also have archival Far Ultraviolet Spectroscopic Explorer (\emph{FUSE}) spectra with high enough S/N to measure \ion{O}{6} absorption. All spectra
have a signal-to-noise ratio S/N$>$7 per resolution element.
Our analysis reveals pervasive low- and high-ion absorption in several phases centered around the LMC both on the sky and in velocity, with radial velocities distinct from the Milky Way. 
The \ion{C}{4} absorption has a high covering fraction of $78\%$ within $25~\mathrm{kpc}$, but between $25~\mathrm{kpc} < \rho_\mathrm{LMC} < 35~\mathrm{kpc}$ the covering fraction is $30$\%. 
Fig. 1 
shows two maps of the Magellanic System superimposed on 21-cm \ion{H}{1} emission maps of neutral hydrogen \cite{hi4pi}, 
with our sightline locations color-coded by \ion{C}{4} column density and mean velocity.
We limit our analysis of the high-ion absorption to components with the following properties:
(1) high-ion column densities not explained by photoionization; 
(2) velocities $v_\mathrm{LSR} > 150$~\kms\ to select Magellanic gas and avoid Milky Way contamination \cite{deBoer1990,Wakker1997,Lehner2011}; 
(3) velocities not associated with known intermediate- and high-velocity clouds \cite{Wakker1997,Putman2012}.

\begin{figure}[!h]
    \centering
    \includegraphics[width=1.0\textwidth]{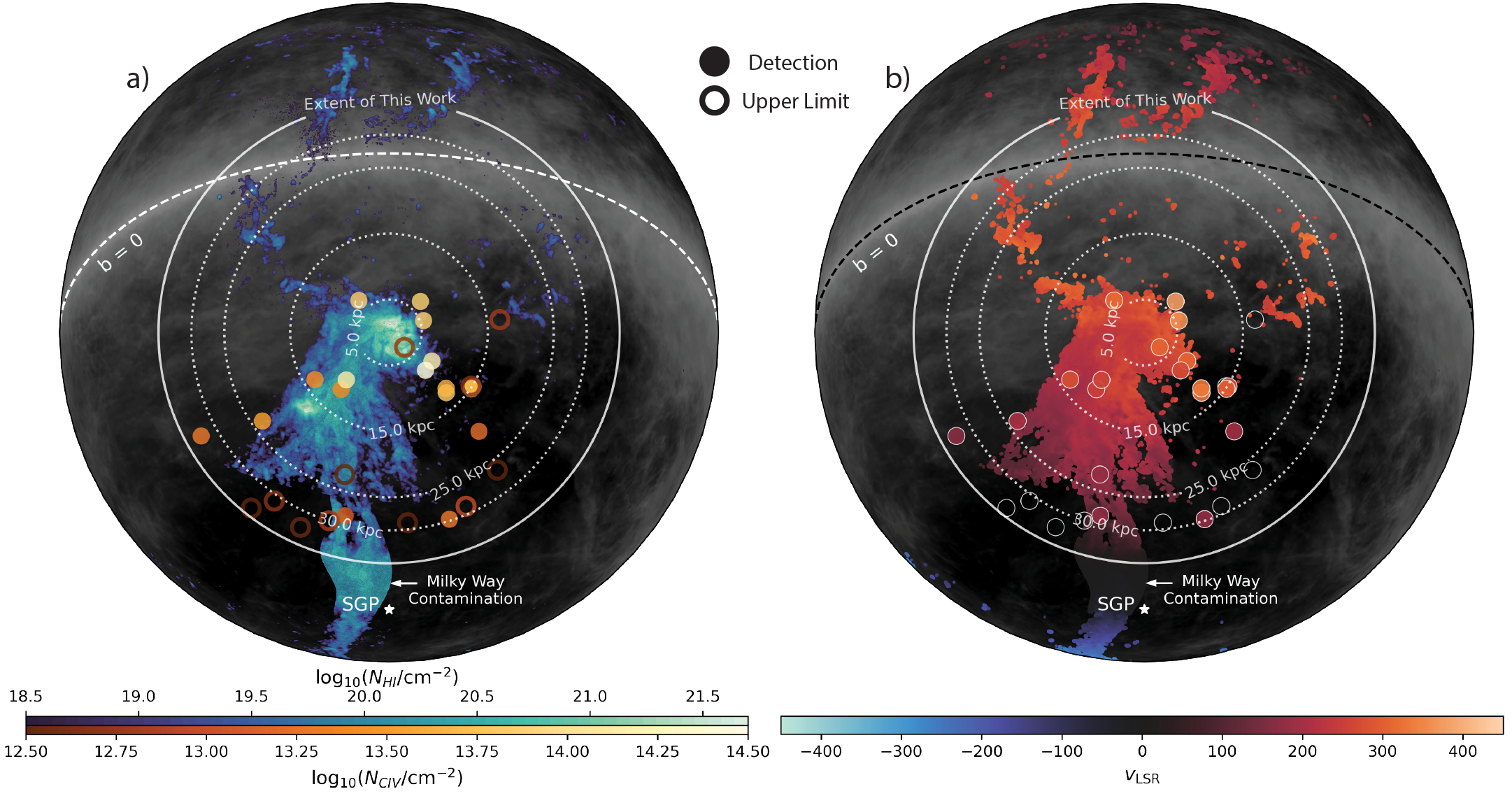}
    \caption{{\bf The Magellanic system in an orthographic projection}. 
    The maps are centered on the LMC and color-coded by column-density (a) and velocity (b). 
    Magellanic 21-cm \ion{H}{1} emission is shown
    in a blue-scale integrated in velocities to encompass the Magellanic System \cite{Nidever2010} with colored symbols showing \emph{HST}/COS sightlines color-coded by
    \ion{C}{4} column densities.
    Upper limits are shown using open symbols. 
    The grey background shows Galactic 21-cm \ion{H}{1} emission from HI4PI \cite{hi4pi} integrated between $-75<v_\mathrm{LSR}<+75$~\kms. 
    Panel b) shows the mean velocity of \ion{H}{1} with our sightlines color-coded by mean \ion{C}{4} absorption velocity. 
    Dotted circles mark the LMC impact parameter and the south galactic pole (SGP) is marked with a white star.}
    \label{fig:1}
\end{figure}

Photoionization accounts for the low-ion absorbers (singly and doubly ionized species) with temperatures $\log_{10}(T_e/K) = 4.02^{+0.07}_{-0.04}$, densities $\log_{10}(n_e/\mathrm{cm}^{-3}) = -1.4\pm0.3$, and line-of-sight cloud sizes $\log_{10}(L/\mathrm{kpc}) = -1.5^{+0.6}_{-0.4}$. 
The high-ion absorbers have column densities too high to be explained by photoionization,
and instead are well explained by equilibrium or non-equilibrium (time-dependent) collisional ionization models\cite{Gnat2007}. 
The \ion{C}{4}/\ion{Si}{4} column-density ratio primarily yields a temperature of $T\sim10^{4.9}~\mathrm{K}$, but solutions as low as $T\sim10^{4.3}~\mathrm{K}$ are possible in non-equilibrium conditions for certain metallicities. 
However, the measured ratios of \ion{O}{6}/\ion{C}{4} and \ion{O}{6}/\ion{Si}{4}
yield higher temperatures.
Our ionization modeling and the component kinematics instead suggest that the \ion{O}{6} ions are tracing a distinct and hotter phase of gas near $\sim10^{5.5}$~K, where \ion{O}{6} peaks in fractional abundance in collisional ionization \cite{Gnat2007}.

We show the relation between the \ion{C}{4}, \ion{Si}{4}, and \ion{O}{6} column densities in the Magellanic CGM and the LMC impact parameter in Fig. 2/
A significant declining profile is observed for \ion{C}{4} and \ion{Si}{4}, indicating that the gas content falls off with radius, a characteristic signature of a diffuse CGM \cite{Tumlinson2017}. 
Sightlines at small impact parameters of $\rho_\mathrm{LMC} < 7~\mathrm{kpc}$ tend to show a deficit of collisionally ionized high-ions
compared to sightlines at $7~\mathrm{kpc}< \rho_\mathrm{LMC} < 12~\mathrm{kpc}$. 
These inner-CGM absorbers are more susceptible to photoionization and winds, due to their close proximity to the LMC. 
When only considering the absorbers at $\rho_\mathrm{LMC} > 7~\mathrm{kpc}$, the significance of the anti-correlation between $N$(\ion{Si}{4}) and $\rho_\mathrm{LMC}$ becomes stronger.
A similar trend is seen with \ion{O}{6}, but is more uncertain owing to the small sample size.

\begin{figure}[!ht]
    \centering
    \includegraphics{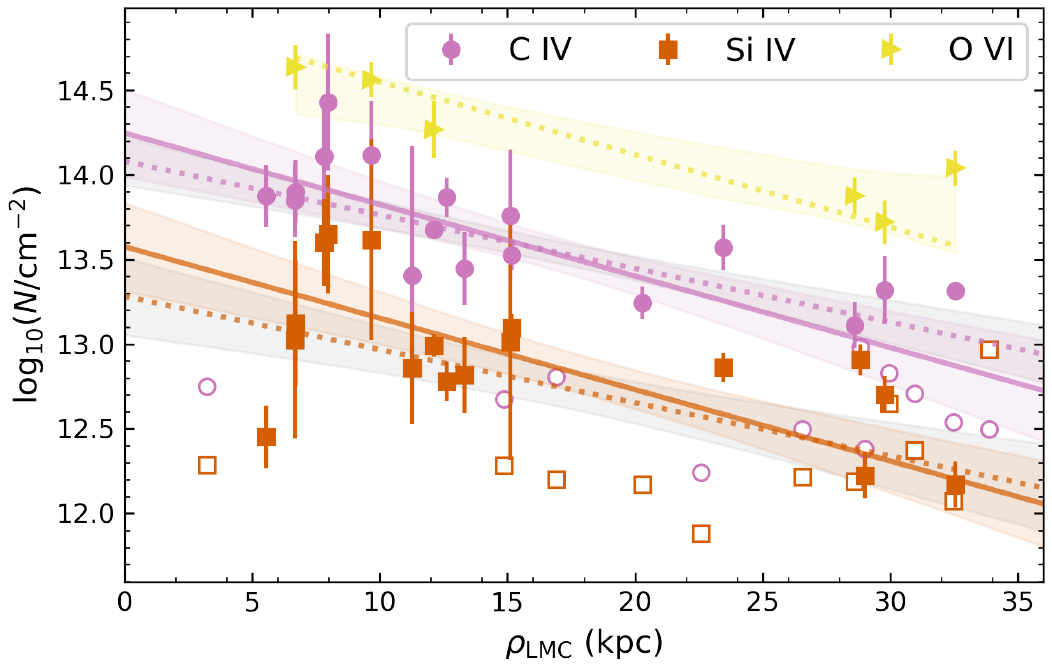}
    \caption{{\bf Declining radial profile of high-ion absorption}. Sum total of \ion{O}{6} (yellow triangles), \ion{C}{4} (pink circles) and \ion{Si}{4} (orange squares) column density of Magellanic velocity absorbers that are not photoionized in each sightline as a function of LMC impact parameter, $\rho_\mathrm{LMC}$, with open symbols marking upper limits. The best-fit line and $68\%$ confidence intervals for all data (dotted line) and only data at  $\rho_\mathrm{LMC}>7$~kpc (solid line) are also shown (see Methods). Error bars represent standard deviations.}
    \label{fig:2}
\end{figure}

For the $\sim10^4$~K low-ion CGM phase, the relation between the modeled ionized hydrogen column density ($N_{\rm \ion{H}{2}}$)
and the impact parameter ($\rho_\mathrm{LMC}$) is shown in panel a) of Fig. 3.
The ionized hydrogen column densities for the $\sim10^{4.9}$~K CGM phase traced by \ion{C}{4} and \ion{Si}{4} and derived from both an equilibrium and non-equilibrium collisional ionization model \cite{Gnat2007} are shown in panel b). 
Both the low- and high-ion gas show similar radial profiles, and all but one sightline where high-ion absorption is observed also show low-ion absorption.
A linear-regression model is used to find the ionized hydrogen mass of these two phases within $\rho_\mathrm{LMC} < 35~\mathrm{kpc}$, using a measured metallicity of $\mathrm{[Z/H]}$ = $-$0.67 for the photoionized gas and an assumed metallicity of $\mathrm{[Z/H]}$ = $-$1 for the high-ion gas.
We find a total ionized hydrogen mass in the $\sim10^4$~K phase of $\log_{10}(M_\mathrm{\ion{H}{2},photo}/M_\odot) = 8.7^{+0.2}_{-0.1}$.
For the $\sim10^{4.9}$~K phase, we find a similar total ionized hydrogen mass of $\log_{10}(M_\mathrm{\ion{H}{2},\,Eq}/M_\odot) = 8.5\pm 0.1$ from a equilibrium model, while an isochoric non-equilibrium model results in $\log_{10}(M_\mathrm{\ion{H}{2},\,Non-Eq}/M_\odot) = 8.6\pm 0.1$.
If this high-ion gas was instead at a lower temperature of $\sim10^{4.3}$~K as is possible in the isochoric model, the mass would increase by an order of magnitude.

\begin{figure}[!h]
    \centering
    \includegraphics[width=1.0\textwidth]{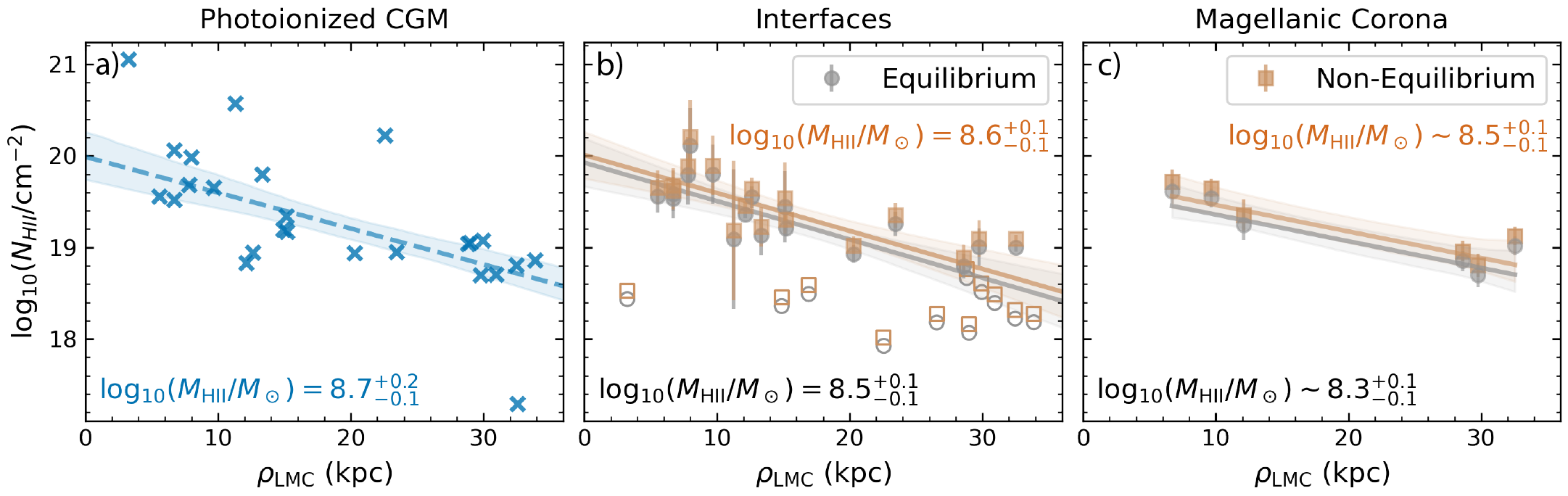}
    \caption{{\bf Total ionized hydrogen profile from ionization models}. Ionized hydrogen column density as a function of LMC impact parameter, $\rho_\mathrm{LMC}$, for each of three gas phases at $\sim10^4$~K (photoionized; low ions; a), $\sim10^{4.9}$~K (interfaces; \ion{C}{4} and \ion{Si}{4}; b), and $\sim10^{5.5}$~K (Corona; \ion{O}{6}; c). 
    The photoionized results are from \emph{Cloudy} models (blue X's). 
    Collisional ionization equilibrium (grey circles) and non-equilibrium isochoric models (brown squares) are used for the $\sim10^{4.9}$, and $\sim10^{5.5}$~K gas based on \ion{C}{4} and \ion{O}{6} measurements, respectively \cite{Gnat2007}.
    Linear regression fits and $68\%$ confidence intervals are shown in the same color lines, with the $\sim10^{4.9}$~K fit made only to data at $\rho_{\mathrm LMC}>7$~kpc, and the $\sim10^{4}$~K and $\sim10^{5.5}$~K fits made to all data. The slope of the $\sim10^{5.5}$~K is only considered within the bounds of our observations. Error bars represent standard deviations. }
    \label{fig:3}
\end{figure}

For the $\sim10^{5.5}$~K CGM phase traced by \ion{O}{6}, we again assume a metallicity of $\mathrm{[Z/H]}$ = $-$1 and 
derive an ionized hydrogen mass within the range of LMC impact parameters where \ion{O}{6} is observed to be $\log_{10}(M_\mathrm{\ion{H}{2}}/M_\odot) \sim 8.3\pm0.1$ for the equilibrium and  $\log_{10}(M_\mathrm{\ion{H}{2}}/M_\odot) \sim 8.5\pm0.1$ for the isochoric non-equilibrium model. 
Since we do not expect any of our high-ion gas phases to maintain collisional ionization equilibrium, the non-equilibrium case is more likely. 
Combined across phases, we estimate a total ionized Magellanic CGM gas mass of $\log_{10}(M_\mathrm{\ion{H}{2},CGM}/M_\odot) \approx 9.1 \pm 0.2$.

We consider three possible explanations for the 
\ion{C}{4}- and \ion{Si}{4}-bearing gas around the LMC. It could exist in either 
a) a diffuse Magellanic Corona at $\sim10^{5}$~K,
b) turbulent or conductive interfaces \cite{Borkowski1990,Kwak2015} between cool, low-ion gas clouds and a hotter diffuse Magellanic Corona at $\sim10^{5.5}$~K, or
c) turbulent or conductive interfaces between cool, low-ion gas clouds and a
hot ($\sim10^6$ K) gaseous Milky Way halo. 
Fig. 4
shows a cartoon schematic of these three scenarios.
When considering all our observations, we conclude that the \ion{C}{4} exists in the interfaces between cooler $\sim10^{4}$~K clouds and a $\sim10^{5.5}$~K Magellanic Corona (model b). 
This model explains the high-ion radial profile, because there are more cool clouds closer in to the LMC, each with interfaces tracking the kinematics of low-ions. 
It also explains the presence of \ion{O}{6}, which shows velocity offsets and directly traces the $\sim10^{5.5}$\ K Corona, but some of which may exist in interfaces.
The line widths are also consistent, with \ion{Si}{4} line widths that are broader than the \ion{Si}{2} line widths as expected with interfaces (though a difference is not confirmed at high significance between \ion{C}{4} and \ion{C}{2} line widths).
The alternative models are less favored because they either result in a thermally unstable Corona with no explanation for \ion{O}{6} (model a) or cannot explain the radial profile dependent on distance from the LMC, not distance from the Milky Way (model c).

\begin{figure}[!h]
    \centering
    \includegraphics[width=1.0\textwidth]{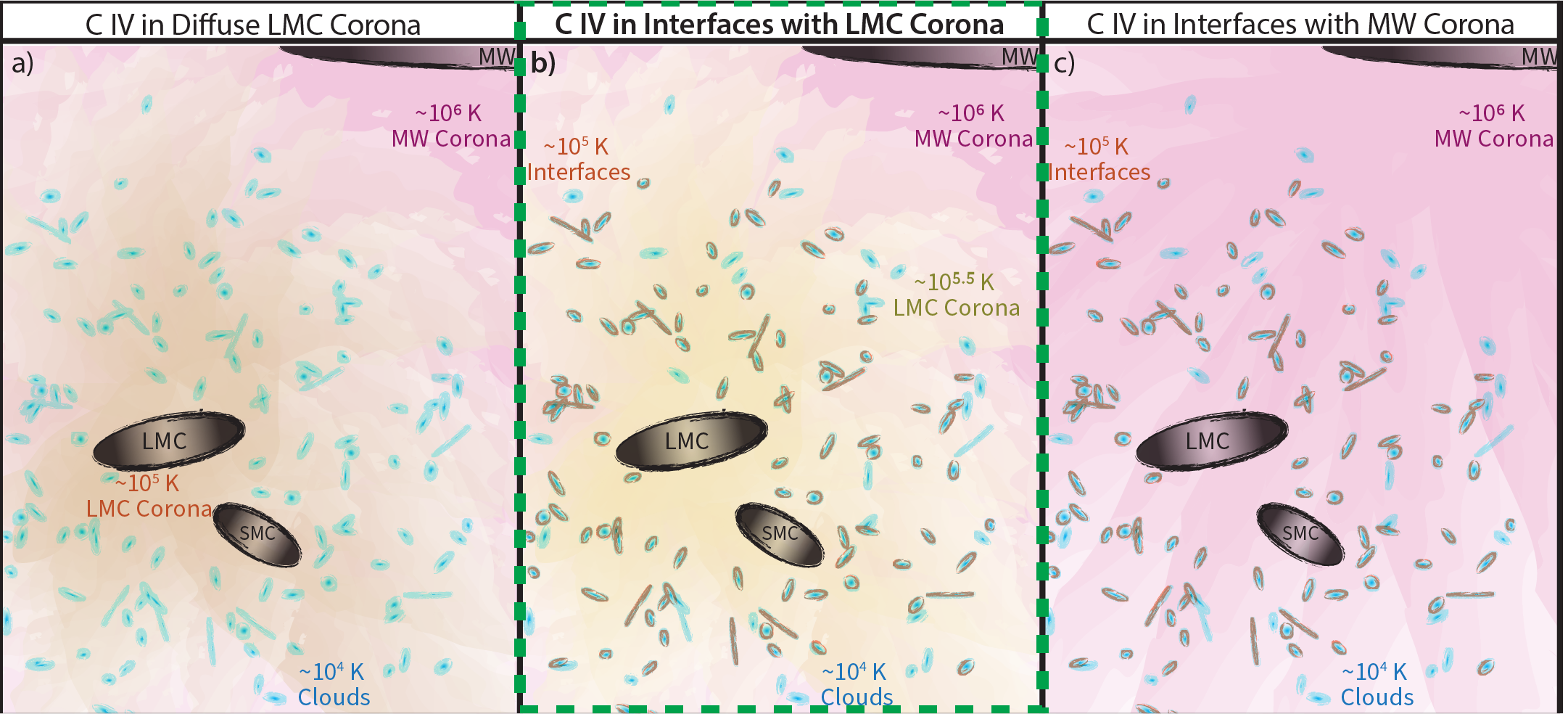}
    \caption{{\bf Summary of three possible \ion{C}{4} CGM models}. Our three possible models for \ion{C}{4} in 
    a) a LMC Corona,
    b) interfaces with an LMC Corona,
    c) interfaces with a Milky Way Corona.
    Cartoon renditions show $\sim 10^4$ K, $\sim 10^5$ K, $\sim 10^{5.5}$ K, and $\sim 10^6$ K gas in blue, orange, yellow, and pink hues, respectively.
    The green dashed-line outlines model b) which best explains our observations. }
    \label{fig:4}
\end{figure}

Several sightlines in our sample have been studied as part of a survey of the Magellanic Stream\cite{Fox2014}, where some high-ion absorption may be attributed to photoionization associated with a Seyfert flare in the Milky Way \cite{JBH2019, Fox2020} or to a shock cascade \cite{JBH2007}. 
However, the modeled Seyfert flare only affects gas within a relatively small ionization cone outside of our sample.
Gas associated with the Stream may contribute to the absorption measured in several other sightlines in our sample, but overall the tidally-stripped Stream gas is less likely to be a major contributor to the Magellanic CGM since it cannot easily explain the radial profile seen in both low- and high-ions. 
The radial profile is also seen in sightlines away from the Stream, further supporting that our observations are not biased by the tidally-stripped gas.
This radial profile appears truncated when compared to those seen in a previous survey of the CGM of $43$ low mass, $z<0.1$ dwarf galaxies (COS Dwarfs) \cite{Bordoloi2014} or the profile seen around the more massive Andromeda Galaxy, M31 \cite{Lehner2020}, though large uncertainties in virial radius estimates make such a comparison difficult (see Extended Data Figure 2, panel b). 

While some sightlines in our sample pass closer to the SMC than the LMC, the common history of the two galaxies should result in a single enveloping Magellanic Corona dominated by the LMC, which is $\sim 10$ times more massive \cite{Besla2012}. While in isolation, SMC-mass galaxies are not massive enough to host their own warm coronae, they can host cool gas in their halos \cite{Bordoloi2014}. However, the strong interactions with the LMC and the Milky Way would have strongly disrupted such a cool halo during the SMC's infall, and so 
a single Magellanic Corona dominated by the LMC prevails.

The Magellanic Corona should be detectable via its dispersion measure induced in radio observations of extragalactic fast radio bursts\cite{Prochaska2019}, since it contains a high column density of free electrons. 
The presence of such a pervasive corona around the LMC supports the picture of a hierarchical evolution for the Local Group, in which the LMC and SMC accreted onto the MW as part of a larger system of dwarf galaxies, a Magellanic Group\cite{D'Onghia2008,Nichols2011,Yozin2015}, not in isolation. 
Earlier work has detected the (ultra-faint) galaxies associated with the Magellanic Group \cite{Bechtol2015,Belokurov2016}; our evidence for the Magellanic CGM and Corona suggests we have now detected its gas, an important part of its baryon budget. 
This provides a more complete understanding of the overlapping and co-evolving ecosystems within the Local Group.

{\renewcommand{\markboth}[2]{}
\def\mnras{Mon. Not. R. Astron. Soc.}
\def\nat{Nature}
\def\apj{Astrophys. J.}
\def\apjs{Astrophys. J. Supp.}
\def\apjl{Astrophys. J. Lett.}
\def\aap{Astron. Astrophys.}
\def\araa{Annu. Rev. Astron. Astrophys.}

 \newcommand{\noop}[1]{}
}

\newpage
\renewcommand{\figurename}{Extended Data Figure}
\setcounter{figure}{0}
\renewcommand{\tablename}{Extended Data Table}


\section*{Methods}

This work uses archival \emph{HST}/COS and \emph{FUSE} spectra to show evidence for the Magellanic Corona. Here we describe our data reduction, Voigt profile fitting, and ionization modeling methods.
Throughout this work, all reported values and uncertainties are medians and $68\%$ confidence intervals, unless otherwise specified. 

\subsection*{Impact Parameters and Projection Effects}

Unlike CGM studies of extragalactic systems, this work focuses on the CGM surrounding the LMC at a distance of just $D = 50$ kpc \cite{Pietrzynski2019}. This proximity means that background QSOs at large angular separations from the LMC, $\theta$, correspond to relatively small physical separations. The impact parameter, $\rho_\mathrm{LMC}$, is found using $\rho_\mathrm{LMC} = D \sin\left(\theta\right)$. At 
$\theta>45$\textdegree, this assumption no longer results in realistic impact parameter estimates, and we would require a true 3D model of the location of gas absorbers to calculate a physical separation between gas absorbers and the LMC. Additionally, for sightlines at large $\theta$ it is harder to kinematically distinguish absorption lines from the LMC and the Milky Way. A larger scale understanding of the Magellanic Corona and multiphase CGM would only be possible with more reliance on models and simulations to identify the 3D locations of gas absorbers. To keep this work more focused on the observationally-derived results, our analysis is thus strictly limited to sightlines within $45$\textdegree\ of the LMC, corresponding to $35$ kpc in impact parameter. 

\subsection*{\emph{HST}/COS Observations}

We design our sample to consist of \emph{HST}/COS FUV observations of background quasars using both the G130M and G160M gratings, covering the wavelength ranges $\approx 1150-1450$\,\AA\ and $\approx 1405-1775$\,\AA, respectively. 
The combination of these gratings enable us to examine the following absorption lines: \ion{O}{1} $\lambda$~1302, \ion{N}{1} $\lambda\lambda$~1199, 1200, 1200.7, \ion{C}{2} $\lambda$~1334, \ion{Al}{2} $\lambda$~1670, \ion{Si}{2} $\lambda\lambda$~1260, 1193, 1190, 1526, 1304, \ion{S}{2} $\lambda\lambda$~1250, 1253, 1259, \ion{Fe}{2} $\lambda\lambda$~1608, 1144, \ion{Si}{3} $\lambda$~1206, \ion{C}{4} $\lambda\lambda$~1548, 1550, and \ion{Si}{4} $\lambda\lambda$~1393, 1402. 
The COS spectra are processed following previously developed custom reduction and wavelength calibration methods \cite{Fox2014,Wakker2015}
based on the raw products from the \emph{calcos}\cite{hstcos_handbook} data reduction pipeline.
In order to remove geocoronal airglow contamination in \ion{O}{1} $\lambda$~1302 and \ion{Si}{2} $\lambda$~1304, we use a
second \emph{calcos} reduction of the data, using only observations taken during orbital night-time.

The COS FUV observations have a native pixel size of $2.5~\mathrm{\kms}$ and a spectral resolution (FWHM) of $\approx 20~\mathrm{\kms}$ and $\approx 15~\mathrm{\kms}$ for G130M and G160M spectra, respectively. 
We bin all spectra such that the resulting spectra are Nyquist sampled with two pixels per resolution element.

\subsection*{\emph{FUSE} Observations}

For $15$ sightlines in our sample, archival \emph{FUSE} spectra are also available and analyzed to search for \ion{O}{6} $\lambda$~1031, 1037 
absorption. 
However, only $6$ sightlines had high enough signal-to-noise to make a measurement. 
These wavelengths fall on the \emph{FUSE} LIF1 channel with a spectral resolution of $\sim 20~\mathrm{\kms}$ and native pixel size of $2~\mathrm{\kms}$, which we bin to Nyquist sample with 2 pixels per resolution element. 
These \emph{FUSE} data are reduced and aligned following the customized methods similar to those used for the \emph{HST}/COS spectra \cite{Wakker2003,Wakker2006}.
The \ion{O}{6} $\lambda$1031 may have contamination from molecular $H_2$ absorption at $\lambda$~1032.356, which corresponds to $\sim130$ \kms\ in the \ion{O}{6} frame. 
However, the expected contribution from this contamination is very small because of the high Galactic latitude of the sightlines, and in most cases negligible. 

\subsection*{Absorption-Line Measurements}

We use the open-source Python software, \emph{VoigtFit} \cite{voigtfit}, to perform Voigt profile fitting of the absorption in several ions observed with \emph{HST}/COS with the G130M and G160M gratings. 
This process uses a least-squares optimizer \cite{lmfit} with recent atomic data \cite{Morton2003,Jitrik2004,Cashman2017} and convolves the Voigt profile with an approximate instrumental profile of a Gaussian with FWHM corresponding to the observed grating resolution. 
While this Gaussian approximation of the instrumental profile is not an exact representation, it has been shown to have a nearly negligible effect on fit results for weak high-velocity components \cite{Fox2020}.
For all ion fits, we normalize the spectra using a third-order polynomial fit to the continuum surrounding absorption lines of interest. 
Regions of the absorption spectra that are contaminated by high-redshift absorption components are then flagged to avoid fitting. 

\begin{table}[]
\centering
\resizebox{\textwidth}{!}{%
\begin{tabular}{@{}lllllllllllll@{}}
\toprule
Source Name &
  RA &
  Dec &
  $\rho_\mathrm{LMC}$ &
  $v_\mathrm{\ion{C}{4}}$ &
  $b_\mathrm{\ion{C}{4}}$ &
  $\log_{10}(N_{\rm   \ion{C}{4}}/\mathrm{cm}^{-2})$ &
  $f_\mathrm{PI,\ion{C}{4}}$ &
  $v_{\rm \ion{Si}{4}}$ &
  $b_{\rm \ion{Si}{4}}$ &
  $\log_{10}(N_{\rm   \ion{Si}{4}}/\mathrm{cm}^{-2})$ &
  $f_\mathrm{PI,\ion{Si}{4}}$ &
  $\log_{10}(N_{\rm \ion{C}{4}}/N_{\rm   \ion{Si}{4}})$ \\
 &
  [deg] &
  [deg] &
  [kpc] &
  [\kms] &
  [\kms] &
  &
  &
  [\kms] &
  [\kms] &
  &
  &
  \\ \midrule
RX J0503.1-6634 &
  $75.77$ &
  $-66.56$ &
  3.2 &
  $196 \pm 18$ &
  $45 \pm 27$ &
  $13.20 \pm 0.31$ &
  $6.05 \pm 4.30$ &
  $196 \pm 18$ &
  $42 \pm 20$ &
  $12.87 \pm 0.22$ &
  $5.93 \pm 3.05$ &
  $0.33 \pm 0.38$ \\
RX J0503.1-6634 &
  $75.77$ &
  $-66.56$ &
  3.2 &
  $279 \pm 6$ &
  $45 \pm 12$ &
  $13.80 \pm 0.10$ &
  $1.54 \pm 0.35$ &
  $279 \pm 6$ &
  $39 \pm 15$ &
  $13.14 \pm 0.17$ &
  $3.21 \pm 1.29$ &
  $0.66 \pm 0.20$ \\
RX J0503.1-6634 &
  $75.77$ &
  $-66.56$ &
  3.2 &
  $355 \pm 8$ &
  $23 \pm 13$ &
  $13.30 \pm 0.25$ &
  $4.80 \pm 2.75$ &
  $355 \pm 8$ &
  $40 \pm 11$ &
  $12.97 \pm 0.14$ &
  $4.66 \pm 1.47$ &
  $0.33 \pm 0.28$ \\
RX J0503.1-6634 &
  $75.77$ &
  $-66.56$ &
  3.2 &
  $404 \pm 9$ &
  $24 \pm 11$ &
  $13.29 \pm 0.19$ &
  $4.08 \pm 1.78$ &
  --- &
  --- &
  $<12.29$ &
  --- &
  $>1.01$ \\
PKS0552-640 &
  $88.10$ &
  $-64.04$ &
  5.5 &
  $397 \pm 6$ &
  $40$ &
  $13.34 \pm 0.26$ &
  $<0.01$ &
  $397 \pm 6$ &
  $9$ &
  $12.45 \pm 0.19$ &
  $0.03 \pm 0.01$ &
  $0.89 \pm 0.32$ \\
PKS0552-640 &
  $88.10$ &
  $-64.04$ &
  5.5 &
  $336 \pm 3$ &
  $19 \pm 5$ &
  $13.24 \pm 0.15$ &
  $<0.01$ &
  --- &
  --- &
  $<12.29$ &
  --- &
  $>0.96$ \\
PKS0552-640 &
  $88.10$ &
  $-64.04$ &
  5.5 &
  $262 \pm 4$ &
  $32 \pm 7$ &
  $13.24 \pm 0.07$ &
  $<0.01$ &
  --- &
  --- &
  $<11.99$ &
  --- &
  $>1.25$ \\
PKS0552-640 &
  $88.10$ &
  $-64.04$ &
  5.5 &
  $453 \pm 13$ &
  $40 \pm 14$ &
  $13.26 \pm 0.24$ &
  $<0.01$ &
  --- &
  --- &
  $<12.28$ &
  --- &
  $>0.98$ \\
PKS0637-75 &
  $98.94$ &
  $-75.27$ &
  6.7 &
  $230 \pm 10$ &
  $10$ &
  $12.63 \pm 0.56$ &
  $<0.01$ &
  $230 \pm 10$ &
  $34$ &
  $11.94 \pm 2.26$ &
  $<0.01$ &
  $0.69 \pm 2.33$ \\
PKS0637-75 &
  $98.94$ &
  $-75.27$ &
  6.7 &
  $393 \pm 8$ &
  $14$ &
  $12.91 \pm 0.26$ &
  $<0.01$ &
  --- &
  --- &
  $<12.04$ &
  --- &
  $>0.87$ \\
PKS0637-75 &
  $98.94$ &
  $-75.27$ &
  6.7 &
  $348 \pm 5$ &
  $19 \pm 9$ &
  $13.40 \pm 0.20$ &
  $<0.01$ &
  $348 \pm 5$ &
  $44 \pm 25$ &
  $12.76 \pm 0.26$ &
  $<0.01$ &
  $0.64 \pm 0.33$ \\
PKS0637-75 &
  $98.94$ &
  $-75.27$ &
  6.7 &
  $290 \pm 9$ &
  $37 \pm 19$ &
  $13.52 \pm 0.17$ &
  $<0.01$ &
  $290 \pm 9$ &
  $43$ &
  $12.60 \pm 0.68$ &
  $0.03 \pm 0.04$ &
  $0.92 \pm 0.70$ \\
IRAS Z06229-6434 &
  $98.94$ &
  $-64.61$ &
  6.7 &
  $448 \pm 2$ &
  $27 \pm 4$ &
  $13.53 \pm 0.05$ &
  $<0.01$ &
  $448 \pm 2$ &
  $16 \pm 3$ &
  $12.79 \pm 0.10$ &
  $<0.01$ &
  $0.74 \pm 0.11$ \\
IRAS Z06229-6434 &
  $98.94$ &
  $-64.61$ &
  6.7 &
  $366 \pm 10$ &
  $22$ &
  $13.22 \pm 0.33$ &
  $<0.01$ &
  $366 \pm 10$ &
  $23$ &
  $12.05 \pm 0.91$ &
  $0.02 \pm 0.04$ &
  $1.17 \pm 0.96$ \\
IRAS Z06229-6434 &
  $95.78$ &
  $-64.61$ &
  6.7 &
  $306 \pm 8$ &
  $35 \pm 11$ &
  $13.33 \pm 0.13$ &
  $<0.01$ &
  $306 \pm 8$ &
  $30$ &
  $12.51 \pm 0.45$ &
  $0.04 \pm 0.05$ &
  $0.82 \pm 0.47$ \\
IRAS Z06229-6434 &
  $95.78$ &
  $-64.61$ &
  6.7 &
  --- &
  --- &
  $<12.43$ &
  --- &
  $262 \pm 10$ &
  $10$ &
  $12.02 \pm 0.67$ &
  $0.02 \pm 0.03$ &
  $<0.41$ \\
IRAS Z06229-6434 &
  $95.78$ &
  $-64.61$ &
  6.7 &
  $394 \pm 7$ &
  $11$ &
  $12.86 \pm 0.66$ &
  $<0.01$ &
  $394 \pm 7$ &
  $36$ &
  $12.20 \pm 0.71$ &
  $<0.01$ &
  $0.66 \pm 0.97$ \\
UVQSJ045415.95-611626.6 &
  $73.57$ &
  $-64.61$ &
  7.8 &
  $319 \pm 5$ &
  $19 \pm 11$ &
  $13.50 \pm 0.57$ &
  $<0.01$ &
  $319 \pm 5$ &
  $32 \pm 6$ &
  $13.49 \pm 0.17$ &
  $<0.01$ &
  $0.02 \pm 0.59$ \\
UVQSJ045415.95-611626.6 &
  $73.57$ &
  $-61.27$ &
  7.8 &
  $394 \pm 10$ &
  $28 \pm 14$ &
  $13.12 \pm 0.19$ &
  $<0.01$ &
  --- &
  --- &
  $<12.35$ &
  --- &
  $>0.77$ \\
UVQSJ045415.95-611626.6 &
  $73.57$ &
  $-61.27$ &
  7.8 &
  $216 \pm 5$ &
  $15 \pm 8$ &
  $12.84 \pm 0.15$ &
  $<0.01$ &
  --- &
  --- &
  $<12.39$ &
  --- &
  $>0.46$ \\
UVQSJ045415.95-611626.6 &
  $73.57$ &
  $-61.27$ &
  7.8 &
  $302 \pm 13$ &
  $38 \pm 6$ &
  $13.88 \pm 0.26$ &
  $<0.01$ &
  $302 \pm 13$ &
  $40 \pm 22$ &
  $12.96 \pm 0.55$ &
  $0.02 \pm 0.02$ &
  $0.92 \pm 0.61$ \\
RBS563 &
  $69.62$ &
  $-61.80$ &
  8.0 &
  $311 \pm 13$ &
  $24$ &
  $13.90 \pm 0.59$ &
  $<0.01$ &
  $311 \pm 13$ &
  $30 \pm 16$ &
  $13.39 \pm 0.54$ &
  $<0.01$ &
  $0.51 \pm 0.80$ \\
RBS563 &
  $69.62$ &
  $-61.80$ &
  8.0 &
  $353 \pm 26$ &
  $39 \pm 18$ &
  $14.09 \pm 0.38$ &
  $0.05 \pm 0.04$ &
  $353 \pm 26$ &
  $38 \pm 16$ &
  $13.46 \pm 0.42$ &
  $0.12 \pm 0.12$ &
  $0.63 \pm 0.56$ \\
RBS563 &
  $69.62$ &
  $-61.80$ &
  8.0 &
  $157 \pm 37$ &
  $40$ &
  $13.21 \pm 0.53$ &
  $<0.01$ &
  --- &
  --- &
  $<12.28$ &
  --- &
  $>0.93$ \\
RBS563 &
  $46.90$ &
  $-72.83$ &
  8.0 &
  $253 \pm 5$ &
  $20 \pm 9$ &
  $13.70 \pm 0.14$ &
  $<0.01$ &
  $253 \pm 5$ &
  $27 \pm 6$ &
  $13.30 \pm 0.12$ &
  $<0.01$ &
  $0.40 \pm 0.18$ \\
ESO031-G08 &
  $46.90$ &
  $-72.83$ &
  9.7 &
  $303 \pm 12$ &
  $24 \pm 9$ &
  $13.78 \pm 0.27$ &
  $<0.01$ &
  $303 \pm 12$ &
  $26 \pm 11$ &
  $13.27 \pm 0.44$ &
  $<0.01$ &
  $0.51 \pm 0.52$ \\
ESO031-G08 &
  $46.90$ &
  $-72.83$ &
  9.7 &
  $262 \pm 11$ &
  $19$ &
  $13.64 \pm 0.36$ &
  $<0.01$ &
  $262 \pm 11$ &
  $20$ &
  $12.95 \pm 0.74$ &
  $<0.01$ &
  $0.69 \pm 0.82$ \\
ESO031-G08 &
  $40.79$ &
  $-72.28$ &
  9.7 &
  $214 \pm 18$ &
  $30$ &
  $13.41 \pm 0.38$ &
  $<0.01$ &
  $214 \pm 18$ &
  $27$ &
  $13.13 \pm 0.70$ &
  $0.03 \pm 0.05$ &
  $0.28 \pm 0.80$ \\
UKS0242-724 &
  $40.79$ &
  $-72.28$ &
  11.3 &
  $194 \pm 12$ &
  $24$ &
  $13.21 \pm 1.01$ &
  $7.52 \pm 17.44$ &
  $194 \pm 12$ &
  $50 \pm 8$ &
  $13.04 \pm 0.18$ &
  $5.31 \pm 2.16$ &
  $0.18 \pm 1.02$ \\
UKS0242-724 &
  $40.79$ &
  $-72.28$ &
  11.3 &
  $253 \pm 11$ &
  $50$ &
  $13.67 \pm 0.74$ &
  $0.74 \pm 1.27$ &
  $253 \pm 11$ &
  $31 \pm 17$ &
  $13.10 \pm 0.28$ &
  $1.54 \pm 0.99$ &
  $0.57 \pm 0.79$ \\
UKS0242-724 &
  $66.50$ &
  $-57.20$ &
  11.3 &
  $312 \pm 17$ &
  $36 \pm 12$ &
  $13.41 \pm 0.76$ &
  $<0.01$ &
  $312 \pm 17$ &
  $31 \pm 19$ &
  $12.86 \pm 0.33$ &
  $<0.01$ &
  $0.55 \pm 0.83$ \\
1H0419-577 &
  $66.50$ &
  $-57.20$ &
  12.1 &
  $311 \pm 1$ &
  $24 \pm 2$ &
  $13.55 \pm 0.03$ &
  $<0.01$ &
  $311 \pm 1$ &
  $14 \pm 3$ &
  $12.73 \pm 0.06$ &
  $0.03 \pm 0.00$ &
  $0.82 \pm 0.06$ \\
1H0419-577 &
  $65.22$ &
  $-56.85$ &
  12.1 &
  $356 \pm 3$ &
  $20 \pm 0$ &
  $13.09 \pm 0.09$ &
  $0.02 \pm 0.00$ &
  $356 \pm 3$ &
  $29 \pm 7$ &
  $12.65 \pm 0.08$ &
  $0.05 \pm 0.01$ &
  $0.43 \pm 0.12$ \\
HE0419-5657 &
  $65.22$ &
  $-56.85$ &
  12.6 &
  $300 \pm 4$ &
  $25 \pm 4$ &
  $13.62 \pm 0.10$ &
  $<0.01$ &
  $300 \pm 4$ &
  $32 \pm 11$ &
  $12.78 \pm 0.12$ &
  $0.03 \pm 0.01$ &
  $0.84 \pm 0.15$ \\
HE0419-5657 &
  $30.56$ &
  $-76.33$ &
  12.6 &
  $360 \pm 10$ &
  $35 \pm 7$ &
  $13.50 \pm 0.15$ &
  $<0.01$ &
  --- &
  --- &
  $<12.36$ &
  --- &
  $>1.14$ \\
PKS0202-76 &
  $30.56$ &
  $-76.33$ &
  13.3 &
  $258 \pm 6$ &
  $20 \pm 8$ &
  $13.34 \pm 0.14$ &
  $<0.01$ &
  $258 \pm 6$ &
  $29 \pm 16$ &
  $12.82 \pm 0.23$ &
  $<0.01$ &
  $0.52 \pm 0.27$ \\
PKS0202-76 &
  $30.56$ &
  $-76.33$ &
  13.3 &
  $294 \pm 7$ &
  $16$ &
  $12.80 \pm 0.48$ &
  $<0.01$ &
  --- &
  --- &
  $<12.45$ &
  --- &
  $>0.35$ \\
RBS567 &
  $69.91$ &
  $-53.19$ &
  14.9 &
  --- &
  --- &
  $<12.68$ &
  --- &
  --- &
  --- &
  $<12.28$ &
  --- &
  --- \\
HE0435-5304 &
  $69.21$ &
  $-52.98$ &
  15.1 &
  $296 \pm 16$ &
  $22$ &
  $13.31 \pm 0.45$ &
  $<0.01$ &
  --- &
  --- &
  $<12.29$ &
  --- &
  $>1.01$ \\
HE0435-5304 &
  $69.21$ &
  $-52.98$ &
  15.1 &
  $350 \pm 24$ &
  $33$ &
  $13.34 \pm 0.39$ &
  $<0.01$ &
  $350 \pm 24$ &
  $24$ &
  $12.75 \pm 0.74$ &
  $<0.01$ &
  $0.59 \pm 0.83$ \\
HE0435-5304 &
  $69.21$ &
  $-52.98$ &
  15.1 &
  $250 \pm 5$ &
  $25$ &
  $13.18 \pm 0.32$ &
  $<0.01$ &
  $250 \pm 5$ &
  $9$ &
  $12.67 \pm 0.65$ &
  $<0.01$ &
  $0.51 \pm 0.72$ \\
HE0439-5254 &
  $70.05$ &
  $-52.80$ &
  15.2 &
  $303 \pm 2$ &
  $34 \pm 6$ &
  $13.40 \pm 0.06$ &
  $<0.01$ &
  $303 \pm 2$ &
  $15 \pm 6$ &
  $13.10 \pm 0.08$ &
  $<0.01$ &
  $0.30 \pm 0.10$ \\
HE0439-5254 &
  $70.05$ &
  $-52.80$ &
  15.2 &
  $243 \pm 4$ &
  $15$ &
  $12.92 \pm 0.16$ &
  $<0.01$ &
  --- &
  --- &
  $<12.28$ &
  --- &
  $>0.64$ \\
PKS0558-504 &
  $89.95$ &
  $-50.45$ &
  16.9 &
  --- &
  --- &
  $<12.81$ &
  --- &
  --- &
  --- &
  $<12.20$ &
  --- &
  --- \\
PKS0355-483 &
  $59.34$ &
  $-48.20$ &
  20.3 &
  $203 \pm 7$ &
  $44 \pm 6$ &
  $13.24 \pm 0.10$ &
  $<0.01$ &
  --- &
  --- &
  $<12.17$ &
  --- &
  $>1.07$ \\
FAIRALL9 &
  $20.94$ &
  $-58.81$ &
  22.6 &
  $185 \pm 3$ &
  $13 \pm 8$ &
  $12.87 \pm 0.29$ &
  $5.51 \pm 3.66$ &
  $185 \pm 3$ &
  $24 \pm 5$ &
  $12.79 \pm 0.06$ &
  $3.95 \pm 0.57$ &
  $0.08 \pm 0.29$ \\
RBS1992 &
  $350.46$ &
  $-70.45$ &
  23.4 &
  $244 \pm 8$ &
  $33 \pm 11$ &
  $13.27 \pm 0.13$ &
  $0.02 \pm 0.00$ &
  $236 \pm 4$ &
  $12$ &
  $12.31 \pm 0.14$ &
  $0.09 \pm 0.03$ &
  $0.96 \pm 0.19$ \\
RBS1992 &
  $350.46$ &
  $-70.45$ &
  23.4 &
  $173 \pm 8$ &
  $32 \pm 11$ &
  $13.27 \pm 0.14$ &
  $<0.01$ &
  $186 \pm 3$ &
  $20 \pm 5$ &
  $12.72 \pm 0.06$ &
  $0.05 \pm 0.01$ &
  $0.55 \pm 0.15$ \\
HE0331-4112 &
  $53.28$ &
  $-41.03$ &
  26.6 &
  --- &
  --- &
  $<12.50$ &
  --- &
  --- &
  --- &
  $<12.21$ &
  --- &
  --- \\
RBS144 &
  $15.11$ &
  $-51.23$ &
  28.6 &
  $181 \pm 10$ &
  $44 \pm 7$ &
  $13.11 \pm 0.14$ &
  $<0.01$ &
  --- &
  --- &
  $<12.19$ &
  --- &
  $>0.92$ \\
HE0246-4101 &
  $42.03$ &
  $-40.81$ &
  28.8 &
  --- &
  --- &
  $<12.98$ &
  --- &
  $179 \pm 6$ &
  $25 \pm 10$ &
  $12.91 \pm 0.09$ &
  $<0.01$ &
  $<0.08$ \\
HE0153-4520 &
  $28.81$ &
  $-45.10$ &
  29.0 &
  --- &
  --- &
  $<12.38$ &
  --- &
  $197 \pm 5$ &
  $12$ &
  $12.22 \pm 0.14$ &
  $0.02 \pm 0.01$ &
  $<0.16$ \\
HE0226-4110 &
  $37.06$ &
  $-40.95$ &
  29.8 &
  $203 \pm 3$ &
  $12$ &
  $12.70 \pm 0.36$ &
  $0.02 \pm 0.01$ &
  $203 \pm 3$ &
  $14 \pm 5$ &
  $12.58 \pm 0.07$ &
  $0.03 \pm 0.01$ &
  $0.12 \pm 0.36$ \\
HE0226-4110 &
  $37.06$ &
  $-40.95$ &
  29.8 &
  $160 \pm 6$ &
  $33 \pm 14$ &
  $13.20 \pm 0.15$ &
  $<0.01$ &
  $160 \pm 6$ &
  $11$ &
  $12.09 \pm 0.21$ &
  $0.03 \pm 0.02$ &
  $1.11 \pm 0.26$ \\
HE0038-5114 &
  $37.06$ &
  $-40.95$ &
  30.0 &
  --- &
  --- &
  $<12.83$ &
  --- &
  --- &
  --- &
  $<12.65$ &
  --- &
  --- \\
HE2336-5540 &
  $354.81$ &
  $-55.40$ &
  30.9 &
  --- &
  --- &
  $<12.71$ &
  --- &
  --- &
  --- &
  $<12.37$ &
  --- &
  --- \\
HE0003-5023 &
  $1.43$ &
  $-50.12$ &
  30.9 &
  --- &
  --- &
  $<12.54$ &
  --- &
  --- &
  --- &
  $<12.07$ &
  --- &
  --- \\
IRAS F21325-6237 &
  $324.09$ &
  $-62.40$ &
  32.5 &
  $170 \pm 1$ &
  $14 \pm 2$ &
  $13.32 \pm 0.04$ &
  $<0.01$ &
  $170 \pm 1$ &
  $9 \pm 6$ &
  $12.17 \pm 0.13$ &
  $0.02 \pm 0.01$ &
  $1.15 \pm 0.14$ \\
HE2305-5315 &
  $347.16$ &
  $-52.98$ &
  33.9 &
  --- &
  --- &
  $<12.50$ &
  --- &
  --- &
  --- &
  $<12.97$ &
  --- &
  --- \\\bottomrule
\end{tabular}%
}
\caption{{\bf Voigt profile model parameters for the Magellanic CGM}. Best fit model parameters for \ion{C}{4} and \ion{Si}{4} for components associated with Magellanic CGM with $v_\mathrm{LSR} > 150$ \kms.
For each sightline, the LMC impact parameter $\rho$, centroid velocity $v$, Doppler parameter (linewidth; $b$), column density log$_{10}(N/\mathrm{cm}^{-2})$, and photoionized fraction $f_\mathrm{PI}$ ($f_\mathrm{PI} = N_\mathrm{Cloudy}/N_\mathrm{Obs}$) for both ions are given. 
Components with $f_\mathrm{PI} \leq 0.1$ are considered as not photoionized and shown in Fig. 3.
Linewidths from data with low S/N spectra or that are fixed in the fitting process do not show errors.
Uncertainties in this table correspond to $1\sigma$ standard deviations.
Additional fit results for all ions and sightlines can be viewed on \url{https://github.com/Deech08/HST_MagellanicCorona}.}
\label{tab:fit_results}
\end{table}

We first fit the absorption in all low and intermediate ions (\ion{O}{1}, \ion{N}{1}, \ion{C}{2}, \ion{C}{2}*, \ion{S}{2}, \ion{Si}{2}, \ion{Si}{3}, \ion{Al}{2}, \ion{Fe}{2}) simultaneously, allowing component line centers to be tied across ions when they show general agreement. 
The \ion{C}{2}* line always contaminates the measurement of absorption in \ion{C}{2} at $+250$~\kms. 
When there are blended \ion{C}{2} components at this velocity, we fix the \ion{C}{2}* column density to a constant value of $10^{13.8}$~cm$^{-2}$, based on average measurements from previous work \cite{Lehner2004}, but in these cases the measured \ion{C}{2} columns near $+250$\,\kms\ are not used in our analysis. 
The \ion{Si}{3} $\lambda$~1206 transition is frequently saturated, requiring the linewidths to be tied to match the fit \ion{Si}{2} linewidths. 
A minimum allowed linewidth of $9$\,\kms\ is applied based on the instrumental resolution, and maximum linewidths are only added as a constraint for highly blended components if they are needed to converge to a best fit.

\begin{table}[h]
\centering
\resizebox{\textwidth}{!}{%
\begin{tabular}{@{}lllllll@{}}
\toprule
Source Name & RA & Dec & $\rho_\mathrm{LMC}$ & $v_\mathrm{\ion{O}{6}}$ & $b_\mathrm{\ion{O}{6}}$ & $\log_{10}(N_{\rm \ion{O}{6}}/\mathrm{cm}^{-2})$ \\
                  & [deg]  & [deg]  & [kpc] & [\kms]       & [\kms]       &                  \\ \midrule
IRAS Z06229-6434  & 95.78  & -64.61 & 6.7   & $231 \pm 12$ & $40$         & $14.07 \pm 0.14$ \\
IRAS Z06229-6434  & 95.78  & -64.61 & 6.7   & $352 \pm 9$  & $40$         & $14.30 \pm 0.11$ \\
IRAS Z06229-6434  & 95.78  & -64.61 & 6.7   & $466 \pm 11$ & $36 \pm 16$  & $14.07 \pm 0.16$ \\
ESO031-G08        & 46.90  & -72.83 & 9.7   & $231 \pm 10$ & $35$         & $14.17 \pm 0.11$ \\
ESO031-G08        & 46.90  & -72.83 & 9.7   & $308 \pm 5$  & $25$         & $14.33 \pm 0.10$ \\
1H0419-577        & 66.50  & -57.20 & 12.1  & $302 \pm 7$  & $16 \pm  8$  & $14.03 \pm 0.13$ \\
1H0419-577        & 66.50  & -57.20 & 12.1  & $358 \pm 11$ & $28 \pm  18$ & $13.90 \pm 0.22$ \\
RBS144            & 15.11  & -51.23 & 28.6  & $199 \pm 9$  & $32$         & $13.88 \pm 0.12$ \\
HE0226-4110       & 37.06  & -40.95 & 29.8  & $180 \pm 8$  & $26$         & $13.72 \pm 0.13$ \\
IRAS F21325-6237 & 324.09 & -62.40 & 32.5  & $165 \pm 6$  & $31 \pm 10$  & $14.04 \pm 0.10$ \\ \bottomrule
\end{tabular}%
}
\caption{{\bf Voigt profile model parameters for \ion{O}{6}}. Best fit model parameters for \ion{O}{6} for components associated with Magellanic CGM with $v_\mathrm{LSR} > 150$ \kms. Linewidths without errors are shown when they are fixed in the fitting process.
Additional fit results for all ions and sightlines can be viewed on \url{https://github.com/Deech08/HST_MagellanicCorona}.}
\label{tab:OVI}
\end{table}

\ion{C}{4} and \ion{Si}{4} are then fit simultaneously following the same procedure, but independent of the low-ion results to avoid biasing our analysis, since the high-ion component structure may be different.
\ion{O}{6} absorption from \emph{FUSE} is also fit independently when data are available and a reasonable continuum can be determined. 
If the component structure of the low and high-ions match, they are flagged after the fitting process so that their column densities, linewidths, and line centers can be compared in the subsequent steps. 
Additionally, we calculate upper limits of any transitions where absorption is not seen based on the S/N of the observed spectra \cite{Jenkins1973,Snow2008}. 
Lastly, fit components attributed to the Milky Way or known intermediate- or high-velocity clouds are flagged to avoid contaminating our analysis. 
We note that some contamination from fixed pattern noise persists in our reduced spectra, which may impact our measured column densities and is not accounted for in our estimated errors.

In total, across 28 sightlines, we initially identify $112$ unique velocity components that may be attributed to the Magellanic System.
We then impose a velocity threshold and only consider absorbers at $v_\mathrm{LSR}>150$~\kms\ to avoid contamination from absorbers associated with the Milky Way \cite{Richter2015}. 
The 150~\kms\ velocity threshold was determined using a combination of the observed component velocities and simulations of the Magellanic System \cite{Lucchini2020}; it represents the value that best separates the Galactic and Magellanic components and is consistent with previous kinematic studies of Magellanic absorption \cite{deBoer1990,Lehner2011}.
Additionally, this velocity threshold is supported by dynamical arguments: given the LMC mass, the virial theorem predicts that Magellanic gas has a velocity dispersion of $50$~\kms\ centered on the LMC velocity of $280$~\kms, implying that $95\%$ of Magellanic gas should be within $180$~\kms\ and $380$~\kms.
As a result, our final sample has $52$ unique Magellanic velocity components that are further analyzed based on their kinematics and photoionization modeling. 
The Voigt-profile model parameters for these $52$ \ion{C}{4} and \ion{Si}{4} components are given in Extended Data Table 1
and the $10$ unique Magellanic \ion{O}{6} absorption components are shown in Extended Data Table 2.
Extended Data Figure 1.
shows our measured \ion{C}{4}$\lambda$~1548 and \ion{O}{6}$\lambda$~1031 absorption-line spectra for our sample. 
Panel a) of Extended Data Figure 2
shows the total measured \emph{HST}/COS column densities 
in a number of low and high ions from the Magellanic absorbers at $v_\mathrm{LSR}>150$~\kms\ 
as a function of LMC impact parameter. 
All low-ions show a declining radial profile, similar to the relation shown in the high-ions (Fig. 2). 

A comparison of our observed radial profile to that seen in the COS Dwarfs survey \cite{Bordoloi2014} and M31 \cite{Lehner2020} is shown in panel b) of Extended Data Figure 2.
We normalize impact parameter measurements across these surveys based on the radius enclosing a mean overdensity of 200 times the critical density, $R_{200}$, which is often used as a measure of the virial radius in CGM studies.
In the radial region of overlap between these surveys and our work, the declining profile of the LMC is more concentrated, with a possibly truncated profile.
Because the LMC halo is already within the virial radius of the MW, it is expected to be tidally truncated, hence such a truncated profile is expected.
However, the uncertainties in estimates of $R_{200}$ are estimated to be $50\%$ in the COS Dwarfs and M31 surveys, with the LMC value we use at $R_{200} = 115 \pm 15$ kpc.

\begin{figure}[!h]
    \centering
    \includegraphics[width=1.0\textwidth]{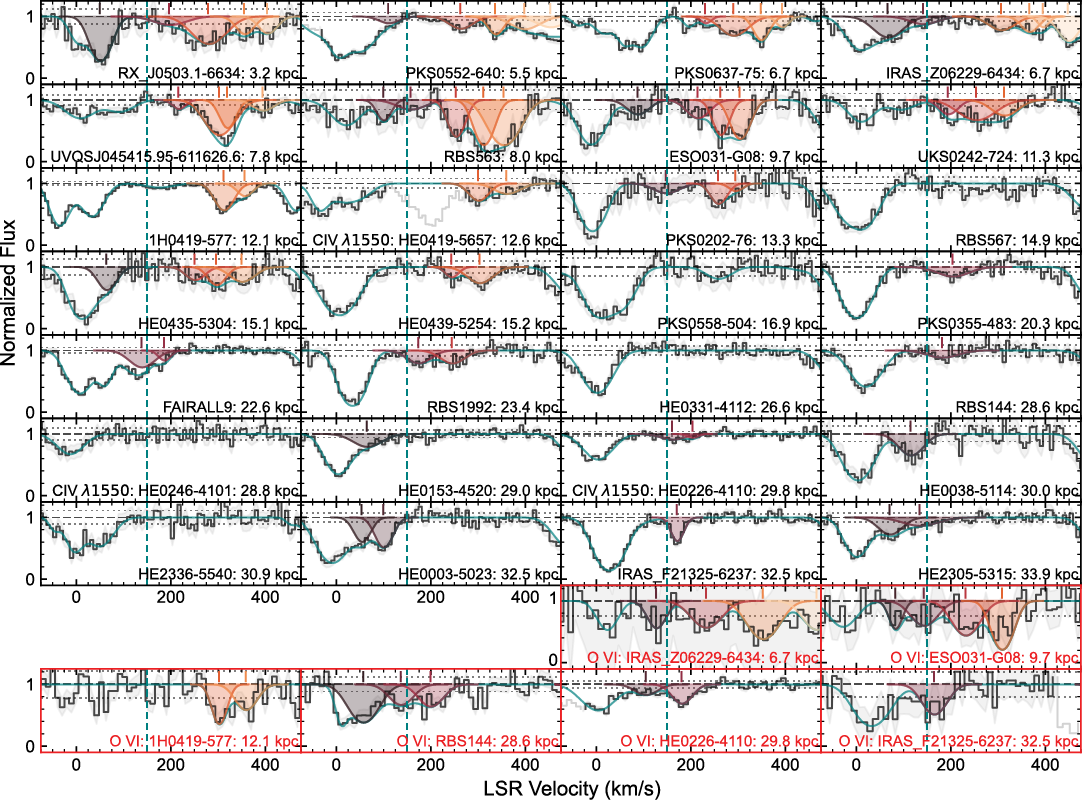}
    \caption{{\bf Sample absorption-line spectra}. Normalized \emph{HST}/COS spectra of the \ion{C}{4}~$\lambda$~1548 (upper panels) and \ion{O}{6}~$\lambda$~1031 (lower panels, red outlines) absorption lines in the LSR velocity frame, ordered by their LMC impact parameters (low to high). 
    The normalized flux is shown in black, with 1$\sigma$ uncertainties shaded in gray around them. 
    The solid teal line shows the full Voigt-profile fit to the \ion{C}{4}~$\lambda$~1548, 1550 and \ion{O}{6}~$\lambda$~1031, 1037 doublets, with individual Magellanic components
    shaded in red-orange hues, 
    corresponding to the same color scheme used in Fig. \ref{fig:1}. 
    Component centers are shown with tick marks.
    The dashed vertical line marks the $150$~\kms\ threshold used in this work. Grayed-out portions of the spectra are contaminated by higher-redshift absorbers and are not considered in our analysis. 
    \ion{C}{4}~$\lambda$~1550 spectra are shown in cases where the \ion{C}{4}~$\lambda$~1548 is highly contaminated.
    Additional fit results for all ions and sightlines can be viewed on \url{https://github.com/Deech08/HST_MagellanicCorona}.}
    \label{fig:S1}
\end{figure}

Our spectra can be accessed publicly on the Barbara A. Mikulski Archive for Space Telescopes (MAST). A full table of our fit parameters, along with summary plots of our best fits can be accessed on GitHub at \url{https://github.com/Deech08/HST_MagellanicCorona}.

\begin{figure}[!h]
    \centering
    \includegraphics[width=1.0\textwidth]{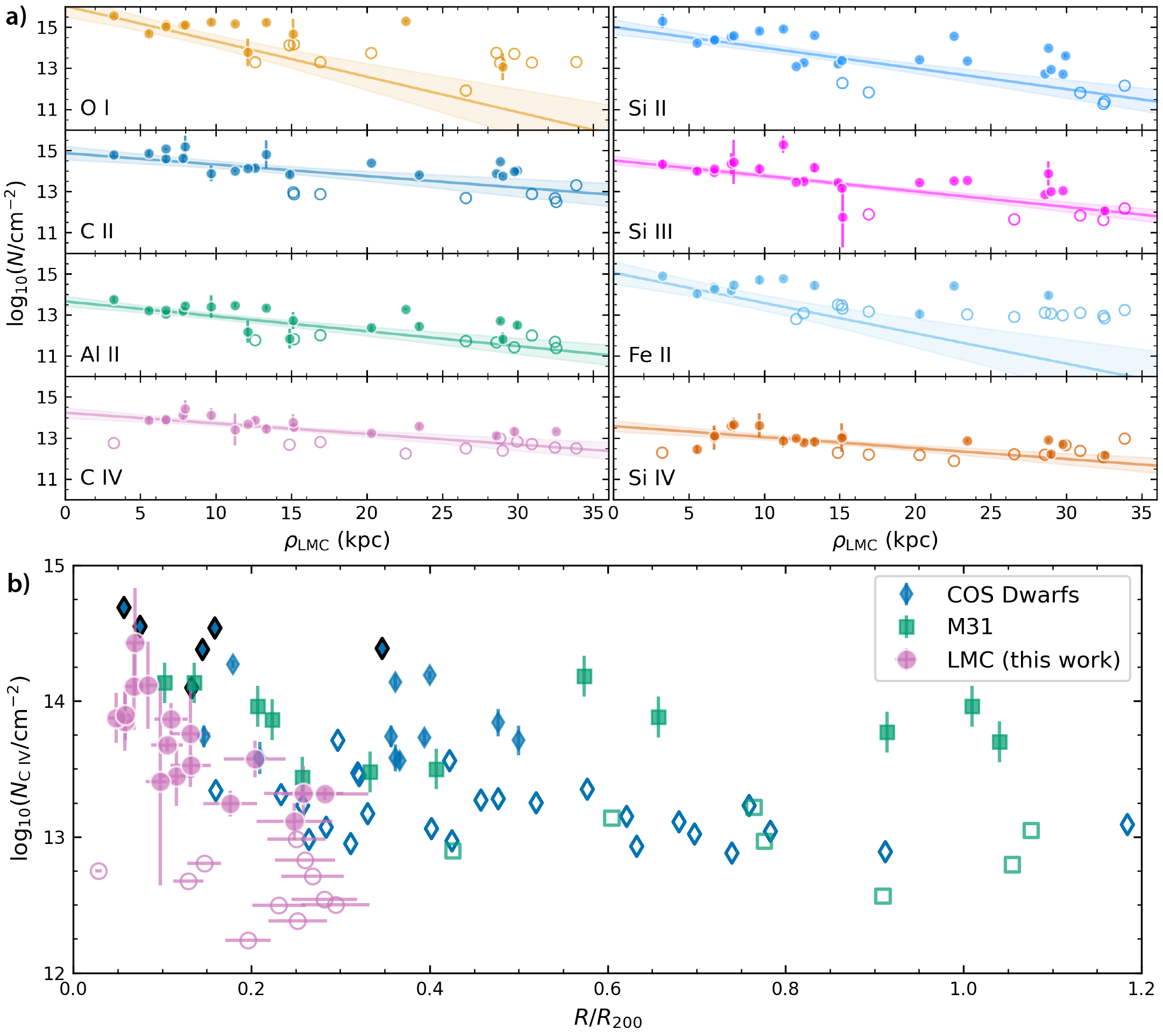}
    \caption{{\bf Radial profiles of ionic column densities}. Measured column densities of Magellanic components at $v_\mathrm{LSR}>150$~\kms\ as a function of LMC impact parameter, $\rho_\mathrm{LMC}$, with open circles marking upper limits in panel a). For each ion, the best-fit line and uncertainties are found using a MCMC analysis with censoring, as in Fig. \ref{fig:3}, for all impact parameter for low-ions and only $\rho_\mathrm{LMC} > 7$ kpc for \ion{C}{4} and \ion{Si}{4}. Each panel corresponds to the ion labeled in the lower left. Panel b) shows a comparison of \ion{C}{4} column densities as a function of impact parameter normalized by the radius enclosing a mean overdensity of 200 times the critical density, $R_{200}$, of the host galaxy with the COS Dwarfs survey \cite{Bordoloi2014} and M31 \cite{Lehner2020}. The LMC data use $R_{200} = 115 \pm 15$ kpc.
    }
    \label{fig:S2}
\end{figure}

\subsection*{Ionization Models}

We use 1D \emph{Cloudy} \cite{Ferland2017} radiative transfer models to simulate the physical conditions of the absorbing gas. 
Our \emph{Cloudy} models require four key inputs in order to run: (1) an external radiation field, (2) the observed column density measurements, (3) a specified stopping condition to reach for convergence, and (4) a gas-phase metallicity. 
All models assume a plane-parallel geometry and constant gas density.

Incident radiation fields in \emph{Cloudy} require a shape and intensity. 
We adopt the Milky Way escaping radiation field model to set the shape of the radiation field, assuming the radiation field from the LMC and SMC have the same spectral shape\cite{JBH1999,Fox2005,Fox2014}. 
The intensity of the radiation field toward each sightline is set by a hydrogen ionizing photon flux $\Phi_{\rm H}$ determined from published ionization models, which includes contributions from the LMC, SMC and Milky Way \cite{JBH2019}.
We reconstruct this model in 3-dimensional space to interpolate an initial value for $\Phi_{\rm H}$ for any specified location. 
In our model, we allow $\Phi_{\rm H}$ to be a free parameter, since a precise distance to the absorbing material is not known.
We also include a constant contribution from an extragalactic UV background \cite{Haardt2012}, and cosmic ray background \cite{Indriolo2007}.

We use \emph{Cloudy}'s built-in \emph{optimize} command to vary our free parameters and find optimal parameters to explain our observed column densities and upper limits \cite{vanHoof1997,Ferland2017}. 
The optimize models use up to three possible free parameters: 
(1) the hydrogen ionizing photon flux, $\Phi_{\rm H}$, described above, 
(2) the total hydrogen number density $n_{\rm H}$, which is the sum of the ionic, atomic, and molecular hydrogen densities of the plasma that is to be modeled, and 
(3) the neutral hydrogen column density ($N_\mathrm{\ion{H}{1}}$) stopping condition.
For sightlines with an \ion{H}{1} or \ion{O}{1} detection, the observed \ion{H}{1} or \ion{O}{1} column density measurement serves as the stopping condition and the model only utilizes the first two free parameters ($\Phi_{\rm H}$, and $n_{\rm H}$).
For sightlines without an \ion{H}{1} or \ion{O}{1}, we utilize all three free parameters ($\Phi_{\rm H}$, and $n_{\rm H}$, $N_\mathrm{\ion{H}{1}}$).
Once \emph{Cloudy}'s \emph{optimize} method has found a possible solution of parameters, we run one final \emph{Cloudy} model at the specified optimal parameters to produce predictions of ion column densities and gas temperatures, including predictions for high-ion (\ion{Si}{4}, \ion{C}{4}, \ion{O}{6}) column densities.
To ensure \emph{Cloudy} does not settle at local minima in the optimization process, we use a broad range of initial densities from $\log_{10}(n_{\rm H}/\mathrm{cm}^{-3})=-3$ to 1, and ionizing fluxes $\Phi_{\rm H}$ in a range of $3$ dex around the model prediction at $D = 50$ kpc, but still find a resulting narrow range of total hydrogen densities ($n_{\rm H}$), ionized gas temperatures $(T_e)$, neutral atomic hydrogen columns $(N_{\ion{H}{1}})$, and ionized-to-neutral atomic hydrogen ratios $N$(\ion{H}{2})/$N$(\ion{H}{1}), across all sightlines and velocity components.
Additionally, we have also run a coarse grid at a larger range of free parameters to help confirm that our solutions are indeed optimal, and not local minima.

While ISM gas-phase metallicities have been measured in the LMC, SMC, Magellanic Bridge \cite{Lehner2008}, and Magellanic Stream \cite{Fox2013,Richter2013,Kumari2015,Howk2017}, the metallicity of 
the Magellanic CGM
is highly uncertain. 
In order to estimate the gas metallicity, we use a sightline in our sample towards HE0226-4110 that overlaps with recently published analysis of \emph{FUSE} spectra to measure neutral hydrogen column densities \cite{French2021}. 
Two absorption components toward this sightline may belong to the Magellanic Corona at $\vlsr = +174~\mathrm{\kms}$ and $+202~\mathrm{\kms}$, providing a measured neutral hydrogen column density to set as a stopping condition in \emph{Cloudy}. 
Unfortunately, there is no detected \ion{O}{1} absorption in either the COS or \emph{FUSE} data, so a metallicity is calculated using a \emph{Cloudy} optimize model (described above), allowing the total hydrogen density, hydrogen ionizing photon flux, and metallicity to vary. 
The \emph{Cloudy} models are optimized based on the measured COS column densities across all available metal ions and any upper limits when absorption is not detected. 
The results for these two components are $\log_{10}(\Phi_{\rm{H}}/\mathrm{photons}~\mathrm{s}^{-1}) = 5.06$, $\log_{10}(n_{\rm{H}}/\mathrm{cm}^{-3}) = -1.58$, and $\mathrm{[Z/H]} = -0.72$ for the $\vlsr = +174~\mathrm{\kms}$ component and $\log_{10}(\Phi_{\rm{H}}/\mathrm{photons}~\mathrm{s}^{-1}) = 4.95$, $\log_{10}(n_{\rm{H}}/\mathrm{cm}^{-3}) = -1.91$, and $\mathrm{[Z/H]} = -0.62$ for the $\vlsr = +202~\mathrm{\kms}$ component. 
Based on these results, we adopt the average $\mathrm{[Z/H]} = -0.67$ as the gas-phase metallicity for photoionized gas.
For hotter gas in interfaces and the corona, we assume a gas-phase metallicity of $\mathrm{[Z/H]} = -1$, since we expect this more primordial gas to be at lower metallicity.

Our optimal set of \emph{Cloudy} models provide predictions for the expected column densities of the high-ions \ion{Si}{4}, \ion{C}{4}, and \ion{O}{6} for a single-phase photoionized gas. 
However, the observed high-ion columns are significantly greater (by orders of magnitude) than the photoionization predictions. 
Across all sightlines and absorption components that may be associated with the Magellanic System, we find that $72\%$ of \ion{Si}{4} and $84\%$ of \ion{C}{4} absorption components are under $10\%$ photoionized. 
We use this $10\%$ ($1~\mathrm{dex}$) threshold to define our sample of  Magellanic absorbers that are not photoionized
(see shaded components in Fig. 2).
These \ion{C}{4} and \ion{Si}{4} absorbers likely arise in interfaces in the range $T = 10^{4.3 - 4.9}~\mathrm{K}$.

\begin{figure}[h]
    \centering
    \includegraphics[width=1.0\textwidth]{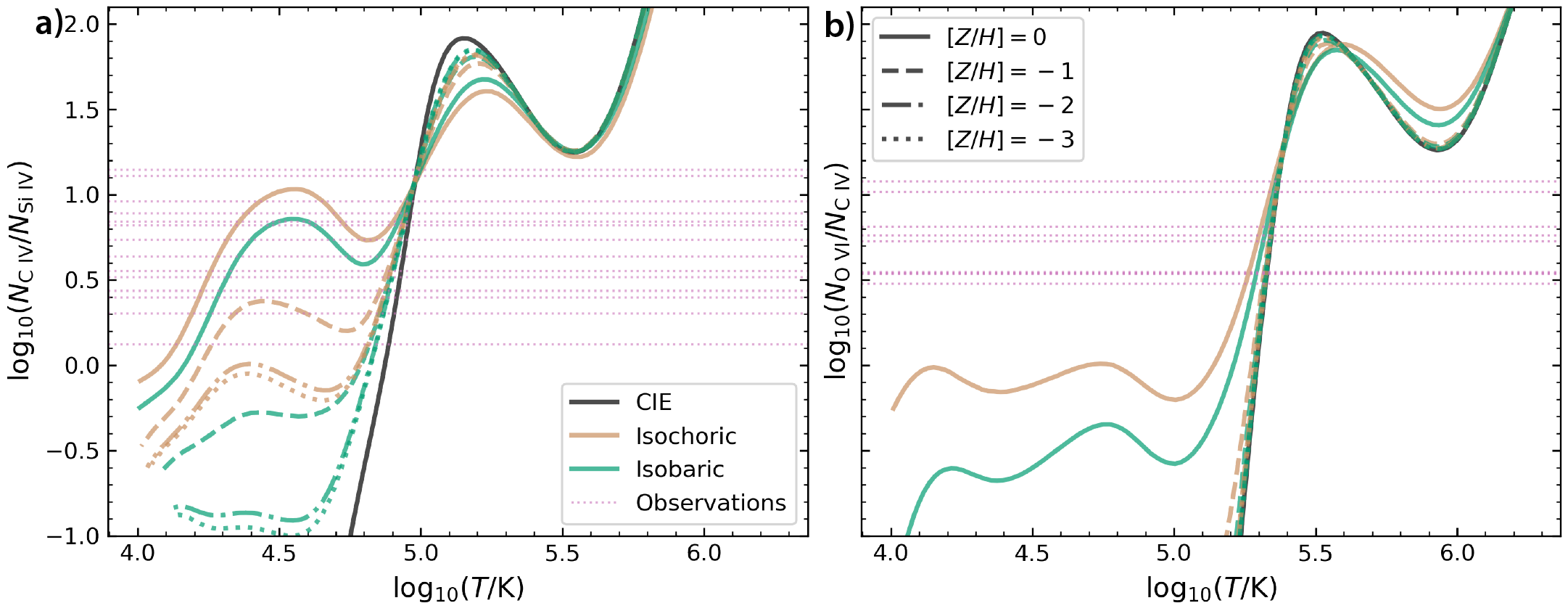}
    \caption{{\bf Collisional ionization model relations}. Predicted column-density ratios as a function of temperature based on the radiative cooling gas models in collisional ionization equilibrium (CIE; black; panel a), time-dependent isochoric cooling (brown; panel b) or time-dependent isobaric cooling (green; panel b) \cite{Gnat2007}. Dotted pink lines show observed values of column density ratios in the Magellanic components. Solid, dash, dash-dot, and dotted lines correspond to metallicities of $[Z/H] = 0$, $-1$, $-2$, and $-3$, respectively.}
    \label{fig:S3}
\end{figure}

The observed triply-ionized Magellanic absorption is well described using either equilibrium or time-dependent 
non-equilibrium collisional ionization models \cite{Gnat2007}. 
In both cases, we can infer an electron temperature based on the ratio of \ion{C}{4} and \ion{Si}{4} column densities, because the close similarity of the \ion{C}{4} and \ion{Si}{4} line profiles indicates the two ions are co-spatial.
The modeled relation of this column-density ratio with temperature for the equilibrium model and for isobaric and isochoric time-dependent models is shown in Extended Data Figure 3
for a range in metallicities. 
The inferred temperature is then used to determine a \ion{C}{4} ionization fraction, from which the total ionized hydrogen (\ion{H}{2}) column density can be calculated, resulting in the measurements shown in Fig. 3
(middle panel).
In total, the temperature distributions of the photoionized and collisionally ionized gas are shown in Extended Data Figure 4
(lower-left panel). In the sightlines where we have measured \ion{O}{6} absorption, we find that the \ion{O}{6} absorbing gas requires a higher temperature than the \ion{C}{4} and \ion{Si}{4} absorbing gas, indicating that the \ion{O}{6} arises in a separate, hotter phase.
While at high metallicity, lower-temperature solutions for our observed column density ratios are possible, this is not the case at the lower metallicities (below 0.1 solar) expected for Magellanic Coronal gas. 

\begin{figure}[h]
    \centering
    \includegraphics[width=1.0\textwidth]{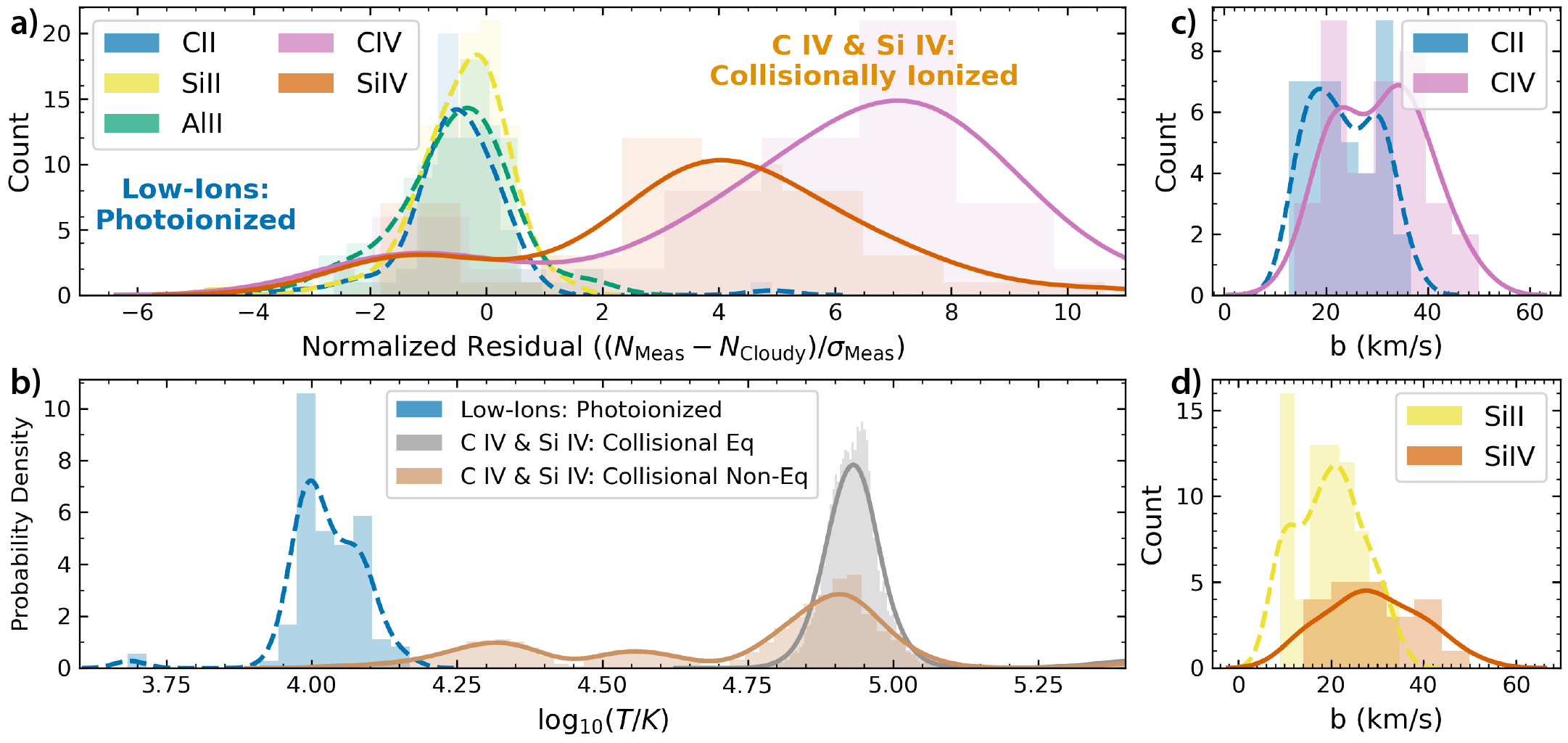}
    \caption{{\bf Evidence for a multi-phase Magellanic CGM}. Panel a) shows histograms and Gaussian kernel density estimates (KDEs) of the residuals between predicted ion column densities from \emph{Cloudy} photoionization models and measured ion column densities, normalized by their standard deviation uncertainties. The histograms are shown for low-ions (\ion{C}{2}, \ion{Si}{2}, and \ion{Al}{2}; dashed lines in blue, yellow, and green, respectively) and high-ions (\ion{C}{4}, \ion{Si}{4}; solid lines in pink, orange).
    Panel b) shows the inferred gas temperatures for the photoionized gas (blue; $\log_{10}(T_e/K) = 4.02^{+0.07}_{-0.04}$) and the collisionally ionized gas under an equilibrium (grey; $\log_{10}(T_e/K) = 4.92^{+0.05}_{-0.02}$) and non-equilibrium isochoric (brown; $\log_{10}(T_e/K) = 4.87^{+0.09}_{-0.06}$; high temperature solution only) model \cite{Gnat2007}. 
    Panel c) and d) show measured linewidths for \ion{C}{2} and \ion{C}{4} absorption (upper; c) and \ion{Si}{2} and \ion{Si}{4} absorption (lower; d).}
    \label{fig:S4}
\end{figure}

We also consider more recent collisional ionization models that include photoionization from an extragalactic background \cite{Gnat2017}.
However, these models do not include the non-isotropic radiation fields necessary for modeling clouds near the Milky Way and LMC, and only offer approximate predictions using a general background radiation field.
Instead, we only consider the two cases of entirely photoionized or entirely collisionally ionized in this work, but note that a full picture will require considering collisional ionization and photoioniation from the Milky Way and Magellanic Clouds together.

\subsection*{Statistical Significance of Results}
Here we describe the statistical tests we used to support our claims of significance. Throughout this work, we adopt a significance threshold p-value of $p=0.05$. 

\begin{figure}[h]
    \centering
    \includegraphics[width=0.7\textwidth]{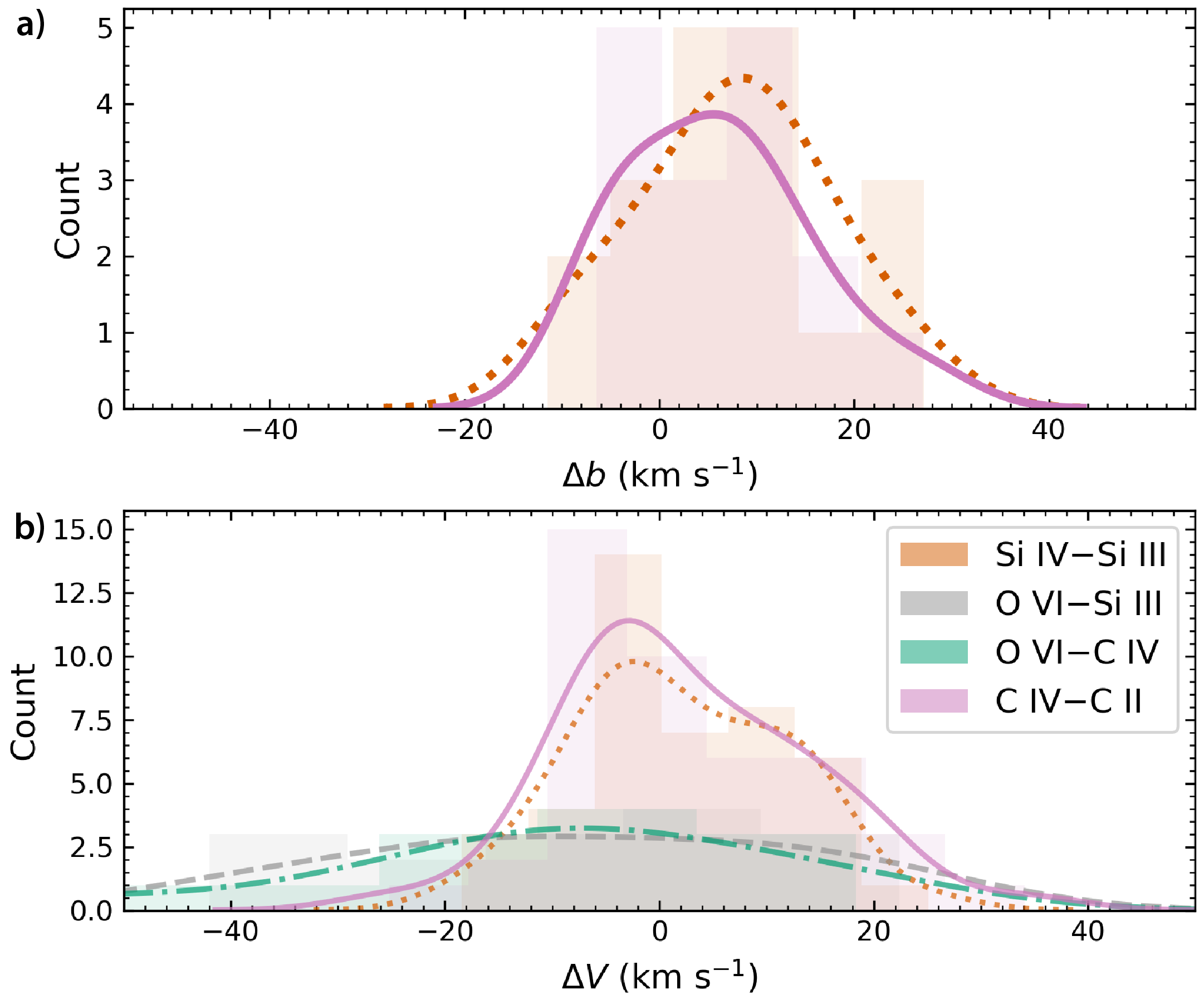}
    \caption{{\bf Kinematic differences of paired high and low ions}. Histograms and Gaussian KDEs of pairs of high $-$ low-ion linewidths (panel a) and velocity centroids (panel b), separated by C ions (pink; solid line) and Si ions (orange; dotted line). The velocity centroid panel also includes differences between \ion{O}{6} and either \ion{Si}{3} (grey; dashed line) or \ion{C}{4} (green; dot-dashed line). The paired linewidths undergo a one-sided Wilcoxon signed-rank test, while the velocity centroids undergo a two-sided test.}
    \label{fig:S5}
\end{figure}

\subsubsection*{Velocity Structure}
In our Voigt-profile fitting process, individual components are initially paired across low- and high-ions based on their approximate centroid velocities. This pairing process is inherently biased as it assumes components across ions are physically tied and results in the lowest possible differences in velocity centroids for our analysis. However, for the low- and high-ions, the velocity structure was qualitatively well matched to one another, with absorption components at similar velocities for both cases. This correspondence is less clear for the \ion{O}{6} absorption line centroids, so matching \ion{O}{6} components in the same manner is much more uncertain. Combined with the relatively low S/N ($\approx$10) and moderate velocity resolution (20 \kms) of our spectra, we are unable to fully resolve all absorption components. 
We therefore find that comparisons of the kinematic properties of low- and high-ions are generally inconclusive. However, the kinematics are still consistent with our primary conclusion that \ion{C}{4} and \ion{Si}{4} arise in the interfaces between cool clouds and a Magellanic Corona,
because in an interface model the velocity structure of the low-ions and high-ions should be linked.
When considering \ion{O}{6}, we calculate the velocity offset from the closest absorption component in other ions (\ion{Si}{3} or \ion{C}{4}), and find the widths of the velocity offset distributions have standard deviations of $\sigma_\mathrm{\ion{O}{6}-\ion{Si}{3}} = 22^{+7}_{-4}$ \kms\ and $\sigma_\mathrm{\ion{O}{6}-\ion{C}{4}} = 22^{+10}_{-6}$ \kms, respectively. This is $7^{+12}_{-8}$ \kms\ greater in width of the distribution of velocity differences between the low-ions and \ion{C}{4} matched in the same manner, supporting the result that \ion{O}{6} exists in a different phase.

\subsubsection*{Linewidths}
We show the paired (matched based on their velocities during the Voigt profile fitting process) differences of component line widths in panel a) of Extended Data Figure 5.
Differences in paired linewidths do not show statistical significance. However, when consider our populations of line width measurements, we do find a statistically significant difference between the line width distributions of singly ionized C and Si, in comparison with triply ionized C and Si (see their distributions in panels c) and d) of Extended Data Figure 4.
The Anderson-Darling statistical test of the null hypothesis that the singly and triply ionized linewidths are drawn from the same underlying population can generally be rejected at the p-value threshold of $0.05$ for both C and Si.
We perform the test on $10,000$ bootstrap samples to account for measurement errors of linewidths. 
The \ion{C}{4} and \ion{C}{2} linewidths return a p-value (with $68\%$ confidence intervals) of $p_\mathrm{C} = 0.008^{+0.08}_{-0.007}$, with $78\%$ of bootstrap samples below our p-value threshold of 0.05.
Similarly, the \ion{Si}{4} and \ion{Si}{2} linewidths return p-values of $p_\mathrm{Si} = 0.001^{+0.01}_{-0.0}$, with $93\%$ of bootstrap samples below our significance threshold.

\begin{figure}[h]
    \centering
    \includegraphics[width=0.7\textwidth]{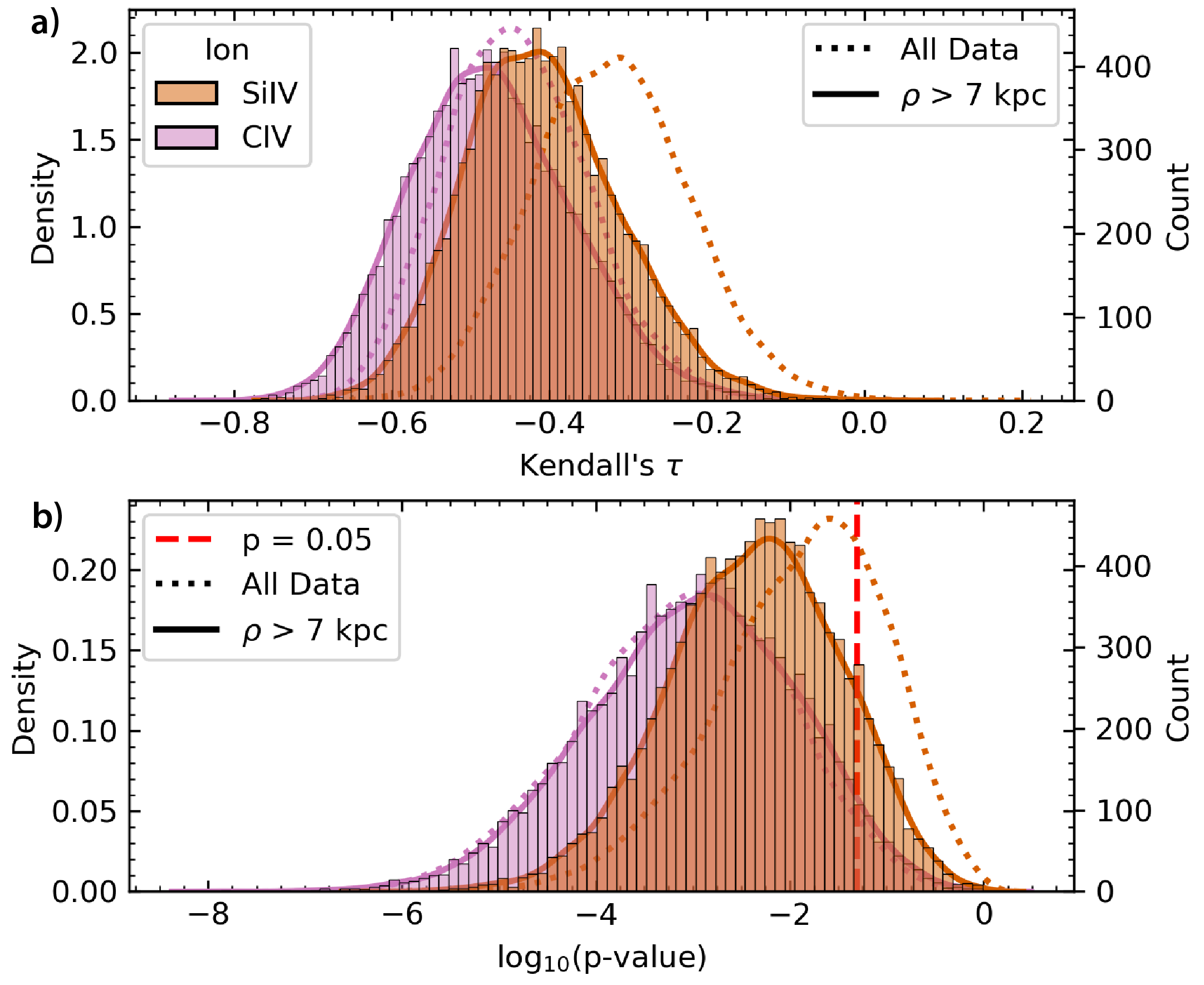}
    \caption{{\bf Statistical significance of radial profile}. Results of a correlation analysis. Kendall's Tau correlation test with censoring \cite{kendalls-tau-censoring} is performed using $10,000$ bootstrap samples, with the resulting distributions of test statistics (panel a) and $p$-values (panel b) shown for the null hypothesis of no correlation. The dotted line shows the Gaussian kernel density estimate for the distribution using all data points, while the solid line and shaded histograms show the result for only data at $\rho_\mathrm{LMC}>7$~kpc.}
    \label{fig:S6}
\end{figure}

\subsubsection*{Declining Radial Profile}

We test the  statistical significance of the anti-correlation between the \ion{C}{4} and \ion{Si}{4} with LMC impact parameter using Kendall's Tau rank correlation coefficient with censoring, which provides a robust measure of the monotonic relationship between two variables \cite{Kendalls-tau,kendalls-tau-censoring}. 
The Magellanic Corona shows a distribution of coefficients that are negative for both \ion{C}{4} and \ion{Si}{4}, with mean values of $\tau = -0.4 \pm 0.1$ and $\tau = -0.3 \pm 0.1$, respectively, as shown in Extended Data Figure 6.
The $p$-values for \ion{C}{4} allow the null hypothesis of no correlation to be rejected at the $0.05$ level for $97\%$ of bootstrap samples, whereas the $p$-values for \ion{Si}{4} can only be rejected for $73\%$ when considering all of our data. 
When only considering the absorbers at $\rho_\mathrm{LMC} > 7~\mathrm{kpc}$, the significance of the \ion{Si}{4} anti-correlation becomes stronger with $p<0.05$ for $89\%$ of $10,000$ bootstrap samples and a mean value of $\tau = -0.4 \pm 0.1$, but the change for \ion{C}{4} is negligible. 
The best-fit lines for the anti-correlation are found using an MCMC analysis with censoring to account for upper limits and measurement errors \cite{Kelly2007}. 
For the \ion{O}{6} measurements, Kendall's Tau rank correlation coefficient is less reliable since we only have 6 data points, and is not conclusive.

\subsection*{Magellanic Corona vs. Tidally Stripped Stream with Interfaces}

Previous simulations have been able to explain much of the ionized gas associated with the Magellanic Stream by tidal stripping, without the presence of a Corona \cite{Wang2019}. 
If this were the case, and the Stream were the dominant source of ionized gas, we would expect to see a stronger correlation of \ion{C}{4} column density as a function of distance from the Magellanic Stream (absolute Magellanic Stream latitude) than as a function of LMC impact parameter. 
We use the partial Spearman rank-order correlation test to assess the strength of the correlation between our measured ion column densities and either the LMC impact parameter or the absolute Magellanic Stream latitude, while removing the effects of the other.
We note that for this test, we are only considering the colissionally ionized \ion{C}{4} and \ion{Si}{4} columns, while considering all the observed columns for low-ions. 
The correlation coefficients and $p$-values of the test with null hypothesis of no correlation are given in Extended Data Table 3.
For most ions, the correlation is significantly stronger with LMC impact parameter, after removing the effects of the absolute Magellanic Stream latitude. However, the partial correlation test for \ion{Fe}{2} is inconclusive and the test for \ion{O}{1} suggests a stronger correlation with absolute Magellanic Stream latitude. These tests are consistent with a Magellanic Corona and CGM origin to the gas absorbers we have measured, with the exception of \ion{O}{1}, which may be more biased towards tracing cooler, tidally stripped gas in the Magellanic Stream.

\begin{table}[]
\centering
\resizebox{\textwidth}{!}{%
\begin{tabular}{@{}|l|l|ll|ll|@{}}
\toprule
Ion & \# of Detections & $r_{N\,\rho_\mathrm{LMC};\left|B_\mathrm{MS}\right|}$ & $p$-value & $r_{N\,\left|B_\mathrm{MS}\right|;\rho_\mathrm{LMC}}$ & $p$-value \\\midrule
\ion{C}{4}  & 17 & $-$0.793 & $<$0.001        & 0.323  & \textbf{0.222} \\
\ion{Si}{4} & 17 & $-$0.524 & 0.037          & 0.224  & \textbf{0.362} \\
\ion{Si}{3} & 22 & $-$0.664 & 0.001          & $-$0.158 & \textbf{0.494} \\
\ion{Si}{2} & 21 & $-$0.583 & 0.007          & $-$0.328 & \textbf{0.159} \\
\ion{C}{2}  & 20 & $-$0.525 & 0.021          & $-$0.076 & \textbf{0.758} \\
\ion{Al}{2} & 18 & $-$0.534 & 0.027          & $-$0.468 & \textbf{0.058} \\
\ion{Fe}{2} & 12 & $-$0.143 & \textbf{0.676} & $-$0.561 & \textbf{0.072} \\
\ion{O}{1}  & 13 & $-$0.117 & \textbf{0.717} & $-$0.693 & 0.012          \\ \bottomrule
\end{tabular}%
}
\caption{{\bf Partial Spearman rank-order correlation tests.} Tests of the relation between ion column density and LMC impact parameter, after removing the effects of absolute Magellanic Stream latitude ($r_{N\,\rho_\mathrm{LMC};\left|B_\mathrm{MS}\right|}$) and between ion column density and absolute Magellanic Stream latitude, after removing the effects of LMC impact parameter ($r_{N\,\left|B_\mathrm{MS}\right|;\rho_\mathrm{LMC}}$). $p$-values for each partial test are shown in boldface if they are greater than the 0.05 threshold, not allowing the null hypothesis of no correlation to be rejected.}
\label{tab:pcor}
\end{table}

In Extended Data Figure 7,
we show our measurements of collisionally ionized \ion{C}{4} columns on a map of the Magellanic System in Magellanic Coordinates, alongside measurements of all \ion{C}{4} absorption from a previous survey of the Magellanic Stream\cite{Fox2014}. 
When considering all \ion{C}{4}, the surface density profile is much more extended along the direction of the Magellanic Stream, but with our adopted velocity threshold and removal of photoionized gas, the radial profile centered on the LMC is apparent, especially when considering sightlines that overlap on our sample and the previous sample. 

\begin{figure}[h!]
    \centering
    \includegraphics[width=1.0\textwidth]{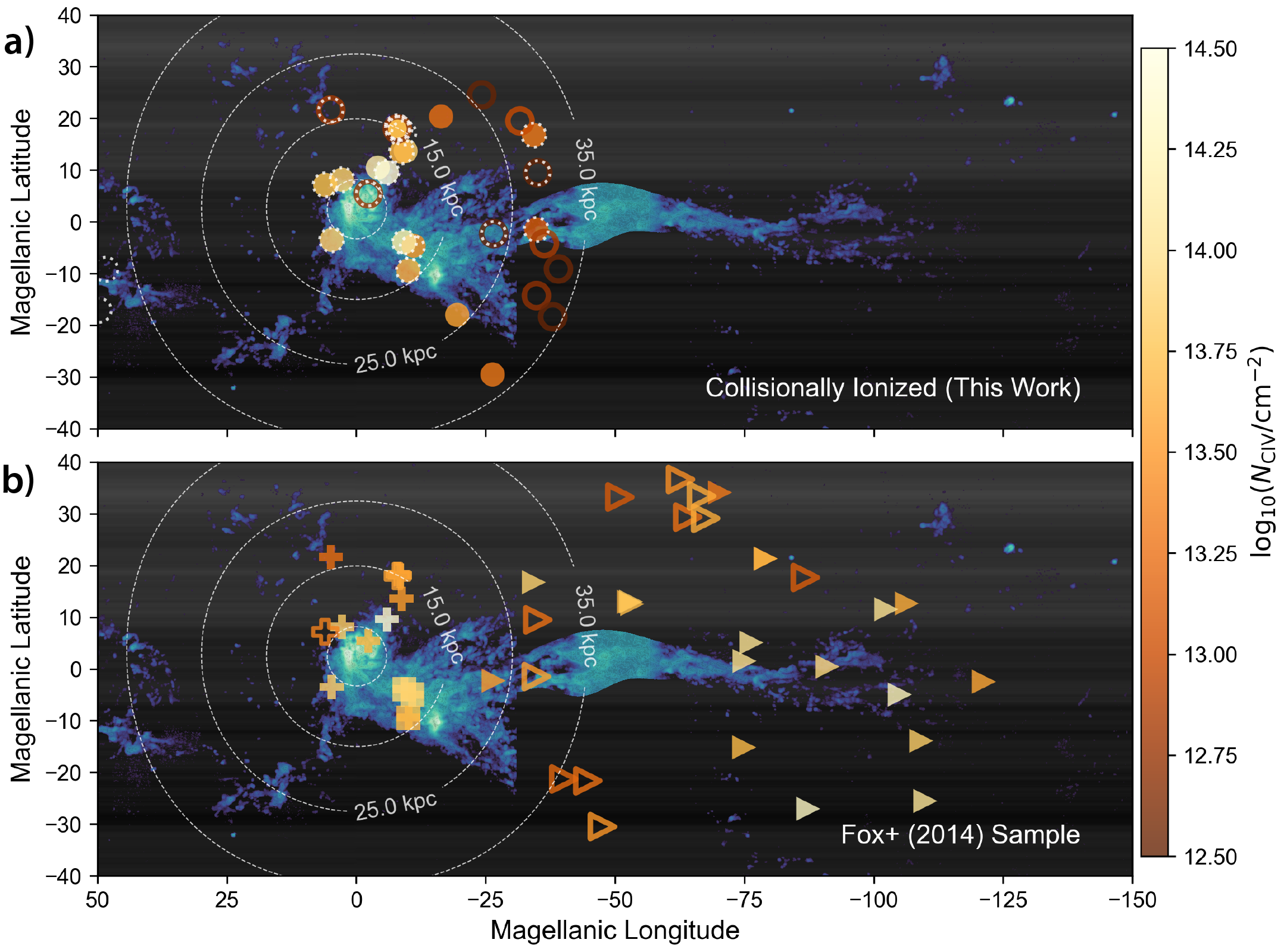}
    \caption{{\bf Magellanic Corona sample vs. Magellanic Stream sample}. Map of measured \ion{C}{4} column density shown in Magellanic Coordinates of our collisionally ionized Magellanic sample (panel a; photoionized phase removed) and previous measurements of all \ion{C}{4} absorption in a survey of the Magellanic Stream\cite{Fox2014} (panel b). Square and triangle symbols in the lower panel mark sightlines that were flagged as belonging to the Magellanic Bridge and Stream, respectively, while other sightlines are marked with a plus symbol. Open symbols show upper limits. Symbols with a dotted white outline in the upper panel denote sightlines that have overlap with the previous survey\cite{Fox2014}, while those without a white outline are unique to this work.}
    \label{fig:S7}
\end{figure}

In this previous work, much of the observed \ion{C}{4} absorption was interpreted to arise from interfaces around the tidally stripped, cooler gas from the LMC with a hot $\sim 10^6$~K Milky Way Corona. 
The basic premise of this conclusion is still valid in our sample, but the strong radial profile centered on the LMC suggests that the hotter gas interacting to form the interfaces should also be centered on the LMC, not the Milky Way. Therefore, a Magellanic Corona at $\sim 10^{5.5}$~K can explain our observed radial profile and the observed \ion{C}{4} absorption.

\subsection*{Mass Estimates}

Our estimates of the mass for each phase of the Magellanic CGM are derived from the relation between ionized hydrogen column density and LMC impact parameter. For each phase ($\sim10^4$~K, $\sim10^{4.9}$~K, $\sim10^{5.5}$~K), a best-fit linear regression model is fit to the ionized hydrogen column as a function of $\rho_\mathrm{LMC}$. Then the ionized hydrogen mass in each phase is calculated using 
\begin{equation}
    M_{\rm{\ion{H}{2}}} = \int_{0~\mathrm{kpc}}^{35~\mathrm{kpc}}~N_{\mathrm{\ion{H}{2}}}\left(\rho_\mathrm{LMC}\right)m_p~2\pi\rho_\mathrm{LMC}~f_\mathrm{cov}~d\rho_\mathrm{LMC}
\end{equation}
where $m_p$ is the proton mass, and $f_\mathrm{cov}$ is the covering fraction. 

For the $\sim10^4$~K gas, 
the ionized hydrogen column density 
in each direction is derived directly from the \emph{Cloudy} models, with a covering fraction $f_\mathrm{cov} = 0.82$, since low-ions are detected at Magellanic velocities in $23/28$ directions in our sample. However, we note that the covering fraction of low-ions tends to decrease as a function of LMC impact parameter, but use a constant covering fraction as an approximation. 

For the $10^{4.9}$~K gas, the total ionized hydrogen column density in each sightline is derived based on the \ion{C}{4} column density and best-fit temperature from the collisional ionization models \cite{Gnat2007} using 
\begin{equation}
    N_{\rm{\ion{H}{2}}} = \frac{N_{\rm{\ion{C}{4}}}}{f_{\rm{\ion{C}{4}}}\left(\rm{Z/H}\right)},
\end{equation} 
where $f_{\rm{\ion{C}{4}}}\equiv \rm{C^{3+}/C}$ is the fraction of triply ionized Carbon at the best-fit temperature, and 
$\left(\rm{Z/H}\right) = 0.21\left(\rm{Z/H}\right)_\odot$ 
is the metallicity. 
For \ion{C}{4}, the covering fraction is set to $f_\mathrm{cov} = 0.78$ for $\rho_\mathrm{LMC} < 30~\mathrm{kpc}$ and $f_\mathrm{cov} = 0.3$ for $\rho_\mathrm{LMC} >= 30~\mathrm{kpc}$ based on the observed detection rate of \ion{C}{4} absorption in our sample. 
The relation between the derived column densities and LMC impact parameter allow for our mass calculations for the $\sim10^4$~K and $\sim10^{4.9}$~K gas to converge, changing by at most $0.1$ dex if instead we integrate out to $500$ kpc. 

For the $\sim10^{5.5}$~K gas, the mass is again found based on the \ion{O}{6} absorption columns in the collisional models, using 
\begin{equation}
    N_{\rm{\ion{H}{2}}} = \frac{N_{\rm{\ion{O}{6}}}}{f_{\rm{\ion{O}{6}}}\left(\rm{Z/H}\right)},
\end{equation} 
and using the same covering fraction correction used for the $\sim10^{4.9}$~K gas. Here, we use the maximal $f_{\rm{\ion{O}{6}}}$ value for each of the collisional models, which peak near $10^{5.5}$~K at $f_{\rm{\ion{O}{6}}} \sim 0.2$.
The best-fit line for this phase does not converge, so integrating the radial profile depends on highly on the radial range considered. Instead, we only integrate between the bounds of our observations ($6.7~\mathrm{kpc} < \rho_\mathrm{LMC} < 32.5~\mathrm{kpc}$) and present an approximate Corona mass for this region only. \\

{\small 
\noindent
\textbf{Data Availability:} \emph{HST}/COS spectra used in this work are publicly available on MAST at \url{https://archive.stsci.edu/}. These archival observations can be found under the following HST Program IDs: 11692, 15163, 12263, 11686, 11520, 12604, 12936, 11541, and 14655. 

\noindent
\textbf{Code Availability:} Voigt-profile fit results, summary fit spectra, and custom code used can be found in our GitHub repository at \url{https://github.com/Deech08/HST_MagellanicCorona}. Additionally, the following software was used in this work. \emph{astropy}\cite{astropy,astropy2}, \emph{calcos}\cite{hstcos_handbook}, \emph{cartopy}\cite{Cartopy}, \emph{lmfit}\cite{lmfit}, \emph{scipy}\cite{scipy}, \emph{voigtfit}\cite{voigtfit}, \emph{Cloudy}\cite{Ferland2017}, \emph{pingouin}\cite{pingouin}

{\renewcommand{\markboth}[2]{}
\def\mnras{Mon. Not. R. Astron. Soc.}
\def\nat{Nature}
\def\apj{Astrophys. J.}
\def\aj{Astron. J.}
\def\apjs{Astrophys. J. Supp.}
\def\apjl{Astrophys. J. Lett.}
\def\aap{Astron. Astrophys.}
\def\araa{Annu. Rev. Astron. Astrophys.}
\def\rmxaa{Rev. Mex. Astron. Astrophys.}

\renewcommand{\refname}{Methods References}

 \newcommand{\noop}[1]{}
}

\noindent
\textbf{Acknowledgments:} This work was funded through HST Archival Program 16363, provided by NASA through a grant from the Space Telescope Science Institute, which is operated by the Association of Universities for Research in Astronomy, Inc., under NASA contract NAS5-26555.
This work uses observations made with the NASA/ESA Hubble Space Telescope, obtained from the Data Archive at the Space Telescope Science Institute, which is operated by the Association of Universities for Research in Astronomy, Inc., under NASA contract NAS5-26555. D.K. is supported by an NSF Astronomy and Astrophysics Postdoctoral Fellowship under award AST-2102490.

\noindent
\textbf{Author Contributions:} D.K. led the investigation, formal analysis, methodology, visualization, and writing. A.J.F. led the project development and management and contributed heavily to the project inception, funding, and writing. E.D.O. led the conceptualization and is Principal Investigator of the \emph{Hubble Space Telescope} grant that funded the research. B.W. led the data curation and contributed heavily to the proposal writing. A.J.F., E.D.O, and B.W. contributed to validation, methodology, and reviewing and editing. D.M.F contributed towards data curation. F.H.C contributed towards methodology and visualization. S.L, F.H.C, D.M.F, J.C.H, and N.L contributed towards validation and reviewing and editing. 

\noindent
\textbf{Competing Interests:} The authors declare no competing interests.

\noindent
\textbf{Additional Information:} Correspondence and requests for materials should be addressed to D.K. (email: dkrishnarao@coloradocollege.edu). 

\noindent
Reprints and permissions information is available at \url{www.nature.com/reprints}.


}


\end{document}